\pdfoutput=1
\documentclass[11pt,twoside,a4paper,cmspaper,final,collab]{cms-tdr}
\def\svnVersion{113f388}\def\svnDate{2024/11/06}\def\cmsCernNoTag{CERN-EP-2023-285}\def\cmsCernDate{\today}\def\cmsMessage{Published in the Journal of High Energy Physics as \href{http://dx.doi.org/10.1007/JHEP10(2024)209}{\doi{10.1007/JHEP10(2024)209}.}}
\begin{document}\cmsNoteHeader{SMP-22-001}

\newlength\cmsFigWidth
\ifthenelse{\boolean{cms@external}}{\setlength\cmsFigWidth{0.85\columnwidth}}{\setlength\cmsFigWidth{0.4\textwidth}}
\newcommand{\ZZ}{\PZ\PZ}
\newcommand{\qq}{\PQq\PAQq}
\newcommand{\pp}{\Pp\Pp\xspace}
\newcommand{\elel}{\ensuremath{2\ell}\xspace}
\newcommand{\elelpr}{\ensuremath{2\ell'}\xspace}
\newcommand{\twoeltwoel}{\ensuremath{\elel\elelpr}\xspace}
\newcommand{\ptvecl}{\ensuremath{\ptvec^{\kern0.2em\ell}\xspace}}

\cmsNoteHeader{SMP-22-001} 
\title{Measurement of differential \texorpdfstring{$\ZZ+\text{jets}$}{ZZ+jets} production cross sections in \texorpdfstring{$\Pp\Pp$}{pp} collisions at \texorpdfstring{$\sqrt{s} = 13\TeV$}{sqrt(s) = 13 TeV}}

\date{\today}

\abstract{
Diboson production in association with jets is studied in the fully leptonic final states, $\pp \to (\PZ/\PGg^*)(\PZ/\PGg^*)+\textrm{jets} \to \twoeltwoel+\text{jets}$, ($\ell,\ell' = \Pe$ or $\PGm$) in proton-proton collisions at a center-of-mass energy of 13\TeV. The data sample corresponds to an integrated luminosity of 138\fbinv collected with the CMS detector at the LHC. Differential distributions and normalized differential cross sections are measured as a function of jet multiplicity, transverse momentum \pt, pseudorapidity $\eta$, invariant mass and $\Delta\eta$ of the highest-\pt and second-highest-\pt jets, and as a function of invariant mass of the four-lepton system for events with various jet multiplicities. These differential cross sections are compared with theoretical predictions that mostly agree with the experimental data. However, in a few regions we observe discrepancies between the predicted and measured values. Further improvement of the predictions is required to describe the $\ZZ$+jets production in the whole phase space.
}

\hypersetup{%
pdfauthor={CMS Collaboration},%
pdftitle={Measurement of the differential ZZ+jets production cross sections in pp collisions at sqrt(s) = 13 TeV},%
pdfsubject={CMS},%
pdfkeywords={CMS, SMP, four leptons, ZZ}}

\maketitle 

\section{Introduction}
\label{sec:introduction}

\begin{figure}[htbp]
 \centering
 \includegraphics[width=0.45\textwidth]{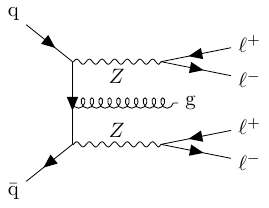}
 \includegraphics[width=0.45\textwidth]{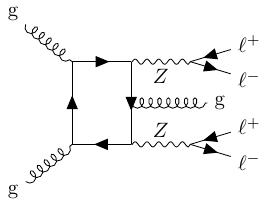}
 \caption{
 Example Feynman diagrams of $\ZZ$ production associated with jets via (left) quark-initiated production and (right) loop-induced
 gluon fusion production.
 }
 \label{fig:Feynman}
 \end{figure}

Measurements of diboson production at the CERN LHC are relevant for precision
studies of the standard model (SM).
In the SM, $\ZZ$ production proceeds mainly through processes represented by the quark-antiquark
$t$- and $u$-channel scattering
diagrams (Fig.~\ref{fig:Feynman} left). In calculations at higher order in quantum chromodynamics (QCD),
gluon-gluon fusion also contributes via
box diagrams involving quark loops (Fig.~\ref{fig:Feynman} right). 
The electroweak (EW) and QCD vertices result in the production of $\PZ$ pairs and of associated jets, and the measurement of this process
is the goal of this analysis.

Previously pairs of on-shell $\PZ$ bosons, produced in the dilepton
mass range 60--120\GeV, were studied by the CMS Collaboration using data sets with
integrated luminosities of 5.1\fbinv at $\sqrt{s} = 7\TeV$~\cite{Chatrchyan:2012sga},
19.6\fbinv at $\sqrt{s} = 8\TeV$~\cite{CMS:2014xja, Khachatryan:2015pba} in
the $\ZZ \to \twoeltwoel$, $\ZZ \to 2\ell2\Pgt$ and $\ZZ \to \elel 2\nu$
decay channels, where $\ell, \ell' = \Pe$ or $\PGm$, and with integrated luminosities of
2.6\fbinv~\cite{Khachatryan:2016txa} and 35.9\fbinv~\cite{Sirunyan:2017zjc} at $\sqrt{s} = 13\TeV$ in 
the $\ZZ \to \twoeltwoel$ decay channel. 
The $\ZZ$ cross section was also measured at $\sqrt{s} = 5.02\TeV$ based on the $\ZZ \to \twoeltwoel$ and $\ZZ \to \elel 2\nu$ decay channels~\cite{CMS:2021pqj}. 
The differential cross sections for $\PZ$ boson pair production in association with jets were measured at $\sqrt{s} = 8$ and $13\TeV$ with integrated luminosities of 
19.7 and 35.9\fbinv, respectively, using the $\ZZ \to \twoeltwoel$ decay channel~\cite{CMS:2018ccg}.
The most recent measurement of ZZ production cross sections with
the full Run 2 data set with an integrated luminosity of 137\fbinv at $\sqrt{s} = 13\TeV$ performed by the CMS Collaboration was published in Ref.~\cite{CMS:2019ZZ},
and results on the EW production of $\ZZ$+2jets were published in Ref.~\cite{CMS:2019EWK}.
All measurements agree with SM predictions. The ATLAS Collaboration published similar
results at $\sqrt{s} = 7$, 8, and 13\TeV~\cite{Aad:2012awa, Aad:2015rka, Aad:2015zqe, Aaboud:2019lxo, Aaboud:2019lgy, Aaboud:2017rwm},
which also agree with the SM. These measurements are important to test predictions 
recently made available at next-to-next-to-leading order (NNLO) in
QCD~\cite{Cascioli:2014yka}. Comparing measurements at the highest collision energies to theoretical predictions
tests the ability of the most advanced higher order QCD and EW calculations to predict the cross sections of complex final states
with jets and multiple vector bosons, and the full Run 2 proton-proton (pp) collision
data at $\sqrt{s} = 13\TeV$ allow diboson measurements at the highest energies
and integrated luminosities to date.

This paper reports a measurement of the four-lepton production
($\pp \to \twoeltwoel$, where $2\ell$ and $2\ell'$ indicate
oppositely charged pairs of electrons or muons, and $\PZ/\PGg^*$ interference is included) in association with jets at $\sqrt{s} = 13\TeV$ using a data set with
an integrated luminosity of $138\fbinv$ recorded in 2016--2018 by the CMS experiment.
Cross sections are reported for the nonresonant production of pairs of $\PZ$ bosons,
$\pp \to \ZZ$, in association with jets, where both $\PZ$ bosons are produced on-shell,
defined as Z bosons with mass in the range 60--120\GeV. The effect of the presence of jets
on the four-lepton mass ($m_{4\ell}$) distribution is also studied with and without 
the on-shell requirement. Differential distributions and cross sections are measured with respect to jet multiplicity ($\mathrm{N}_{\text{jets}}$), transverse momentum \pt, pseudorapidity $\eta$, 
invariant mass and $\Delta\eta$ of the dijet system composed of the highest-\pt and second-highest-\pt jets, and with respect to $m_{4\ell}$ for events with different jet 
multiplicities. The results are compared with predictions of theoretical models. This analysis is an extension to that of Ref.~\cite{CMS:2019ZZ} with a focus on jet variables. The two analyses, Ref.~\cite{CMS:2019ZZ} and this paper, use the same events with a few minor differences: (i) a 0.3\% update in the estimated luminosity of the data set; (ii) the simulation program \MGvATNLO~\cite{MGatNLO} is used
instead of \POWHEG~\cite{Alioli:2010xd,Alioli:2008gx,Nason:2004rx,Frixione:2007vw} as the nominal $\qq \to \ZZ$ sample; 
and (iii) a regularized unfolding method is used instead of a simple matrix inversion. Therefore, the $\ZZ$ fiducial cross section measured in Ref.~\cite{CMS:2019ZZ} is directly valid for this analysis. The results are also compared with recent nNNLO+PS predictions~\cite{Buonocore:2021fnj,Alioli:2021wpn}, where PS is parton shower. 

\section{The CMS detector}

A detailed description
of the CMS detector, together with a definition of the coordinate system used
and the relevant kinematic variables, can be found
in Ref.~\cite{Chatrchyan:2008zzk}.

The central feature of the CMS apparatus is a superconducting solenoid of
6\unit{m} internal diameter, providing a magnetic field of 3.8\unit{T}.
Within the solenoid volume are a silicon pixel and strip
tracker, a lead tungstate crystal electromagnetic calorimeter (ECAL), and a
brass and scintillator hadron calorimeter,
which provide coverage in pseudorapidity $\abs{ \eta } < 1.48 $ in a cylindrical barrel
and $1.48 < \abs{ \eta } < 3.00$ in two endcap regions.
Forward calorimeters extend the
coverage provided by the barrel and
endcap detectors to $\abs {\eta} < 5.0$. Muons are detected in gas-ionization detectors embedded in
the steel flux-return yoke outside the solenoid
in the range $\abs{\eta} < 2.4$, with
detection planes made using three technologies: drift tubes, cathode strip
chambers, and resistive-plate chambers.

Electron momenta are estimated by combining energy measurements in the
ECAL with momentum measurements in the tracker. The momentum resolution for
electrons with $\pt \approx 45\GeV$
from $\PZ \to \Pep \Pem$ decays
ranges from 1.7\% for nonshowering electrons in the barrel region to 4.5\% for
showering electrons in the endcaps~\cite{Khachatryan:2015hwa}.
Matching muons to tracks identified in
the silicon tracker results in a \pt resolution for
muons with $20 <\pt < 100\GeV$ of 1.3--2.0\% in the barrel and better than 6\%
in the endcaps. The \pt resolution in the barrel is better than 10\% for muons
with \pt up to 1\TeV~\cite{Chatrchyan:2012xi,Sirunyan:2018fpa}.

Events of interest are selected using a two-tiered trigger system.
The first level, composed of custom hardware processors, uses information from the calorimeters and muon detectors to select events at a rate of around 100\unit{kHz} 
within a fixed latency of 4\mus~\cite{CMS:2020cmk}. The second level, known as the high-level trigger, consists of a farm of processors running a version of the full event reconstruction software optimized for fast processing, and reduces the event rate to around 1\unit{kHz} before data storage~\cite{Khachatryan:2016bia}.

\section{Data and Monte Carlo samples}
\label{sec:datasets}

The data sample used in this analysis was recorded by the CMS experiment in 2016, 2017, and 2018,
corresponding to 36.3, 41.5, and 59.7\fbinv of integrated luminosities, respectively. 
The details of the luminosity measurement are described in Refs.~\cite{CMS:LUM-17-003,CMS:LUM-17-004,CMS:LUM-18-002}.

The Monte Carlo (MC) simulation used for this analysis can be divided into signal and background samples. 
The $\ZZ$ signal production via quark-antiquark annihilation is simulated at
next-to-leading order (NLO) in QCD with \MGvATNLO v2.4.2~\cite{MGatNLO} 
and \POWHEG~2.0 \cite{Alioli:2010xd,Alioli:2008gx,Nason:2004rx,Frixione:2007vw}. The \MGvATNLO sample is
used as the nominal $\qq \to \ZZ$ sample in reconstruction-level distributions and for unfolding, because this sample is expected to describe data better than \POWHEG since it merges the 0-jet and 1-jet NLO processes, whereas the \POWHEG sample is simulated at NLO accuracy for 0-jet and LO accuracy for 1-jet processes.
The $\Pg\Pg \! \to \! \ZZ$ process is simulated at leading order (LO)
with \MCFM v7.0~\cite{Campbell:2010ff}. The cross sections of these samples are normalized to the
cross sections calculated at NNLO in QCD for
$\qq  \to  \ZZ$ ($K$ factor of 1.1)~\cite{Cascioli:2014yka} and at NLO in QCD for
$\Pg\Pg  \to  \ZZ$ ($K$ factor of 1.7)~\cite{Caola:2015psa}.
The production processes via SM Higgs boson production and decay (specifically $\Pg\Pg \to \PH \to \ZZ$) are simulated with \POWHEG at
NLO. Electroweak $\ZZ$ production in association with two jets is simulated
with \MGvATNLO~\cite{MGatNLO} at LO.
The nominal SM signal predictions are derived from the
\MGvATNLO
$\qq \to \ZZ$ sample, the \MCFM $\Pg\Pg \to \ZZ$ sample, and the
\MGvATNLO EW production sample, which includes vector boson fusion Higgs boson contributions
and their interference with non-Higgs boson EW production, and the \POWHEG $\PH \to \ZZ$ sample.

Simulated events for the irreducible background processes containing four prompt leptons in the
final state, such as $\ttbar\PZ$, $\PW\PW\PZ$, $\PW\PZ\PZ$, and $\PZ\PZ\PZ$, where the last three are 
combined and denoted as VVV, are simulated with \MGvATNLO at LO ($\ttbar\PZ$) and NLO (VVV).

Parton showering, hadronization and fragmentation are simulated in all samples with \PYTHIA 8.226 and 8.230~\cite{Sjostrand:2015}, with parameters set by the CUETP8M1~\cite{Khachatryan:2015pea} (CP5~\cite{CP5}) tune for the 2016 (2017 and 2018) data-taking period,
and the NNPDF3.0 (3.1) set of parton distribution functions, PDFs,~\cite{nnpdf} is used. 

Results are also compared with the very recent nNNLO+PS predictions~\cite{Buonocore:2021fnj,Alioli:2021wpn}, which
consist of NNLO predictions for the quark-initiated production combined with parton
showers using the MiNNLO\textsubscript{PS} method, and NLO predictions for the loop-induced
gluon fusion production matched to parton showers, with event generators for the
two channels implemented in the \POWHEG framework. Spin correlations,
interferences, and off-shell effects are included by calculating the full process $\pp \to \ell^+ \ell^- \ell'^+ \ell'^-$
 and considering all contributions to the four-lepton final state. The contribution mediated by a Higgs boson is included in the gluon fusion production mode.

As part of the nNNLO+PS predictions, the $\qq\to\ZZ$ predictions from the MiNNLO\textsubscript{PS} method are accurate at NNLO for inclusive production and accurate at NLO for $\PZ$+1-jet production. The combination of the two jet
multiplicities does not require any unphysical merging scale~\cite{Monni_2020}. These predictions are expected to be more accurate at high jet multiplicities than: (i) the \POWHEG $\qq  \to  \ZZ$  predictions, which are accurate at NLO in inclusive production;
(ii) the \MGvATNLO predictions,
which are simulated at NLO with the 0- and 1-jet processes, and merged using the FxFx scheme~\cite{Frederix_2012}.  

The detector response is simulated using a detailed description of the CMS detector implemented with the \GEANTfour package~\cite{GEANT}. 
The simulated samples include additional interactions per bunch crossing, referred to as pileup.
Simulated events were weighted so that the pileup distribution reproduces that observed in the data.

\section{Event reconstruction}
\label{sec:eventreconstruction}

Standard CMS reconstruction and identification (ID)
algorithms, referred to as particle-flow (PF)~\cite{Sirunyan:2017ulk}, are used to reconstruct and
identify stable particles arising from collisions---electrons, muons, photons, and charged
and neutral hadrons---by combining the signals from all subdetectors. Electrons
and muons are considered candidates for the reconstruction of ZZ final states
(``signal leptons'') if their $\pt^\ell > 7\,(5)\GeV$ and their
$\abs{\eta^\ell} < 2.5 \,(2.4)$ for electrons (muons).

Signal leptons are required to originate from the primary 
interaction vertex of the event, which is taken to be the vertex corresponding to the hardest scattering in the event, 
evaluated using tracking information alone, as described in Section 9.4.1 of Ref.~\cite{CMS-TDR-15-02}.
The distance of the lepton track origin from the primary vertex
is required to be $<$1\unit{cm} along the beam line, and
$<$0.5\unit{cm} in the transverse plane.
Furthermore, the significance of the three-dimensional impact parameter relative
to the event vertex, $\mathrm{SIP_{3D}}$, is required to satisfy
$\mathrm{SIP_{3D}} \equiv \abs{ \frac{\mathrm{IP}}{\sigma_\mathrm{IP}} } < 4$
for each lepton, where $\mathrm{IP}$ is the distance
of closest approach of the lepton track to the primary vertex
and $\sigma_\mathrm{IP}$ is its associated uncertainty.

Loose and tight ID requirements are defined for each lepton. The tight IDs are used for signal leptons,
whereas the loose IDs are used in control regions to define objects that might be
spuriously identified as a signal lepton. An electron satisfies the loose ID if it satisfies
the \pt, $\eta$, and vertex requirements above. It satisfies the tight ID if it satisfies the loose ID
and the multivariate discriminator described in Ref.~\cite{Sirunyan:2021exp}. A muon satisfies the loose ID if it satisfies the above \pt, $\eta$,
and vertex requirements, and provides a good track-matching between the tracker and the muon detectors. It satisfies the tight ID 
if it satisfies the loose ID, and either is tagged as a muon by the PF algorithm for the years 2016--2017 (satisfies a multivariate discriminator for the year 2018~\cite{CMS:2023dsu}),
or is a high \pt ($>200\GeV$) muon and satisfies a set of requirements on the quality of the associated track.

Signal leptons are required to
 be isolated from other particles in the event. The
relative isolation is defined as
\begin{equation}
        R_\text{Iso} = \Biggl( \sum_\text{charged} \!\!\pt + \max\Bigl[0, \sum_\text{neutral} \!\!\pt
                       + \sum_{\text{photons}} \!\!\pt - \pt^\text{PU} \left(\ell\right)
                       \Bigl]\Biggr) \bigg/ \pt^{\ell}
        \label{eq:iso}
\end{equation}
where the sums run over the \pt of hadrons and photons in a
cone of size $\Delta R \equiv \sqrt{\smash[b]{\left(\Delta\eta\right)^2 + \left(\Delta\phi\right)^2}}$ = 0.3
around the lepton momentum direction, where $\phi$ is the azimuthal angle in radians.
To mitigate the contribution of pileup interactions to the isolation,
charged hadrons are included only if they originate from the
primary vertex~\cite{CMS:2020ebo}. The estimated neutral contribution to the isolation
from pileup, $\pt^\text{PU}(\ell)$,
is defined differently for electrons and muons. For electrons,
$\pt^\text{PU}(\Pe) \equiv \rho \, A_\text{eff}$ where the average \pt
flow density $\rho$ is calculated for each event using a ``jet area'' method~\cite{Cacciari:2007fd}, and is
defined as the median of the $\pt^\text{jet}/A_\text{jet}$ distribution for all pileup jets in the event.
The effective area $ A_\text{eff}$ is the geometric area of the isolation cone projection on the face of the calorimeter multiplied by an $\eta$-dependent correction factor
that accounts for the residual dependence of the isolation value on pileup.
For muons, $\pt^\text{PU}(\mu) \equiv 0.5 \, \sum\limits_{i} \pt^{\text{PU},i}$, where $i$
runs over the momenta of the charged hadron PF candidates originating from pileup vertices, and
the factor of 0.5 corrects for the ratio of charged to neutral particles in the isolation cone.
For the years 2016--2017, muons are considered isolated if their
$R_\text{Iso}< 0.35$, whereas for 2018 and for electrons the isolation requirement is included in the multivariate discriminator used for the selection.

The efficiencies for the reconstruction, identification, and isolation of
signal leptons are measured in data and simulation using a tag-and-probe
technique~\cite{CMS:2011aa} based on inclusive samples of $\PZ$ boson events,
with an additional sample of $\JPsi$ events for low-\pt muons.
The measurements are performed in bins of $\PT^{\ell} $ and $ \abs{\eta^\ell}$, where for electrons the supercluster $\eta$ is used.
The electron selection efficiency in the ECAL barrel (endcaps) varies from
${\sim} 85 \,(77)\%$ at $\PT^{\Pe} \approx 10\GeV$,
to ${\sim} 95 \,(89)\%$ for $\PT^{\Pe} \geq 20\GeV$,
and is ${\sim}85\%$ in the barrel-endcap transition region.
The muons are reconstructed and identified with efficiencies above ${\sim}98\%$
within $\abs{\eta^{\PGm}} < 2.4$. The ratio between the data and simulation
efficiencies in each \pt-$\abs{\eta}$ bin is applied as a correction factor to leptons in simulated events. If the correction factor for a given
lepton is $f_\text{eff}^\ell(\pt{}^\ell,\eta^\ell)$, the efficiency
correction for each event is
$\prod\limits_{\ell} f_\text{eff}^\ell(\pt{}^\ell,\eta^\ell)$, where the
product index runs over the four leptons of the $\ZZ$ candidate.

Jets are reconstructed based on PF candidates, 
rejecting the charged hadrons associated to a pileup vertex,
 with the anti-\kt clustering algorithm~\cite{Cacciari:2008gp,Cacciari:2011ma} using a distance parameter $R = 0.4$. 
To reduce the instrumental background, tight identification criteria based on the multiplicities and energy fractions carried by charged and neutral hadrons are imposed on jets~\cite{CMS-PAS-JME-16-003}. 
Jets from pileup are rejected using pileup jet identification criteria based on the compatibility of the associated tracks with the primary vertex, when inside the tracker acceptance, and on the jet topology~\cite{CMS:2020ebo}. Jet energy corrections are applied to the reconstructed jets~\cite{1748-0221-6-11-P11002, Khachatryan:2016kdb}. 

\section{Event selection}
\label{sec:eventselection}

The data samples used in this analysis are selected by the trigger system that requires the presence of a pair of loosely
isolated leptons or a triplet of leptons, with minimum-\pt thresholds for leptons depending on the lepton combination.
Further triggers include a set of single-electron and single-muon triggers, and
triggers on
leptons of different flavors. The trigger efficiency within
the acceptance is greater than 98\%. 

Events are required to have at least four leptons.
Each event should contain at least one lepton with $\pt > 20\GeV$, two leptons with  $\pt > 10\GeV$, and four leptons with $\pt > 7 \,(5)\GeV$ for electrons (muons). All leptons must pass
the ``tight'' lepton identification and isolation
requirements described in Section~\ref{sec:eventreconstruction}.

The leptons are required to be separated by $\Delta R (\ell_1, \ell_2 ) > 0.02$,
and electrons are required to be separated from muons by
$\Delta R (\Pe, \PGm ) > 0.05$, to remove spurious
``ghost'' leptons arising from ambiguities in track reconstruction.
Lepton pairs originating from hadronic decays are removed by requiring
that all oppositely charged lepton pairs in the $\ZZ$ candidate
have $m_{\ell_1 \ell_2} > 4\GeV$
regardless of lepton flavor.

$\PZ$ candidates are built from two oppositely-charged leptons of the same flavor.
The pair is retained if it satisfies $4 < m_{\ell^+\ell^-} < 120\GeV$.
All possible four-lepton candidates
in an event are then considered. For each $\ZZ$ candidate, the dilepton pair with invariant mass
closest to the nominal $\PZ$ mass $m_{\PZ}$ (91.1876\GeV~\cite{PDG2022}) is designated $\PZ_1$, and the other is designated $\PZ_2$.
The event is kept if
$40 < m_{\PZ_1} < 120\GeV$ and $4 < m_{\PZ_2} < 120\GeV$.

In the case of multiple $\ZZ$ candidates satisfying all requirements,
the ambiguity is resolved by selecting the candidate where $m_{\PZ_1}$
is closest to the nominal $\PZ$ mass. If more than one lepton combination is still possible,
the $\PZ_2$ candidate is chosen as the one that maximizes the scalar \pt sum of the leptons.

In this analysis, the jets are required to have a $\pt > 30\GeV$ and $\abs{\eta} < 4.7$. 
In addition, jets are required to be well separated from any isolated lepton by requiring $\Delta R(\text{jet,lepton}) > 0.4$, where the lepton here
satisfies all the tight requirements except for the lower \pt requirement ($>5\GeV$ instead of $>7\GeV$ for electrons, 
$>3\GeV$ instead of $>5\GeV$ for muons) and a relaxed $\mathrm{SIP_{3D}}$ requirement ($<$10 instead of $<$4 for electrons).

\section{Background estimates}
\label{backgrounds}

The requirement of four well-reconstructed and isolated lepton candidates strongly suppresses any
background; therefore, this analysis has very low background contributions, dominated by
$\PZ$ boson and $\PW\PZ$ diboson production in association with
jets, and \ttbar production.

In a small fraction of cases, particles from jet fragmentation satisfy both lepton identification and
isolation criteria, and thus are misidentified as signal leptons. This background is estimated using control data samples.
The probability for jets to be misidentified and selected as leptons 
is measured from a sample of
$\PZ$+$\ell_\text{candidate}$ events, where $\PZ$ denotes a pair of
oppositely charged, same-flavor leptons that pass the selection requirements and
satisfy $\abs{ m_{\ell^+\ell^-} - m_{\PZ}} < 7\GeV$. Each event in this sample must have exactly one
additional lepton candidate $\ell_\text{candidate}$ that satisfies the loose identification requirements with
no isolation requirements applied. The misidentification probability for
each lepton flavor, measured in bins of \pt and $\eta$ of the $\ell_\text{candidate}$,
is defined as the ratio between the number of candidates that pass the final
isolation and identification requirements to the total number of candidates in the sample.
The number of $\PZ$+$\ell_\text{candidate}$ events is corrected for the contamination
from $\PW\PZ$ production and for $\ZZ$ events in which one lepton is not
reconstructed. These events have a third genuine, isolated lepton that must be excluded
from the misidentification probability calculation. The $\PW\PZ$ contamination is suppressed by requiring the missing
transverse momentum $\ptmiss < 25\GeV$. The $\ptmiss$ is defined
as the magnitude of the missing transverse momentum vector $\ptvecmiss$,
the projection onto the plane transverse to the beams of the negative
vector momentum sum of all reconstructed PF candidates in the event,
corrected for the jet energy scale (JES). 
The transverse mass, calculated as
$\mT \equiv \sqrt{\smash[b]{(\pt^\ell + \ptmiss )^2 -(\ptvecl + \ptvecmiss )^2}}$, is required to be $<$30\GeV. The residual contribution
of $\PW\PZ$ and $\ZZ$ events, which can be up to a few percent of the
$\ell_\text{candidate}$ events passing all selection criteria, is estimated
from simulation and subtracted.

Two control samples are
used to estimate the number of background events
in the signal region. Both are defined as samples that contain events with
a dilepton candidate satisfying all requirements (as $\PZ_1$) and
two additional lepton candidates $\ell^{+}\ell^{-}$.
In one control sample, enriched in $\PW\PZ$+jets events, one
$\ell$ candidate is required to satisfy the tight
identification and isolation criteria and the other must fail this selection 
and instead satisfy only the loose requirements; in the other control sample, enriched in
$\PZ$+jets events, both $\ell$
candidates must satisfy the loose criteria, but fail the full criteria.
The additional leptons must have opposite charges and the same
flavor ($\Pe^{\pm}\Pe^{\mp}$ and $\PGm^{\pm}\PGm^{\mp}$).
The expected number of background events in the
signal region, denoted ``$\PZ$+X'' in the figures, is obtained
by scaling the number of observed $\PZ_1+\ell^{+}\ell^{-}$ events
by the misidentification probability for each lepton failing the selection.
The procedure is described in more detail in Ref.~\cite{Chatrchyan:2013mxa}.

In addition to this reducible background, which contributes to approximately 1--2\% of the expected $\ZZ \to 2\ell2\ell'$ event yield, 
the yields from the $\ttbar\PZ$ and VVV processes 
with four prompt leptons are estimated from simulated samples
to be around 1.0--1.5\% of the expected $\ZZ \to 2\ell2\ell'$ yield.

\section{Unfolding and systematic uncertainties}
\label{sec:unfoldingAndSyst}
To obtain differential cross sections normalized to the $\ZZ$ fiducial cross sections (for the on-shell $\ZZ$ region and for the full four-lepton invariant mass range 
as defined by the kinematic requirements) and compare CMS data to
theoretical predictions, the data are ``unfolded'' to remove detector
resolution, efficiency, and acceptance effects. For each distribution to be
unfolded, a response matrix is obtained from simulated signal samples.
The response matrix represents the correlation map between the distributions obtained after the full detector simulation,
reconstruction, and selection, and the generated distributions they originate
from. It is used in unfolding to obtain true physical distributions from observed data. 
The data are unfolded using the
iterative D'Agostini's method~\cite{DAgostini:1994fjx} including correction for background contributions, with the \texttt{RooUnfold} toolkit as described in Ref.~\cite{Adye:2011gm},
 and compared with
the theoretical predictions from \MGvATNLO $\qq \to \ZZ$ and \POWHEG $\qq \to \ZZ$, 
where {\MCFM} $\Pg\Pg \to \ZZ$, \POWHEG $\PH \to \ZZ$, and \MGvATNLO EW ZZ predictions are also added to these two sets of predictions.
The unfolded results are also compared with the nNNLO+PS predictions.

The measured on-shell $\ZZ$ fiducial cross section from Ref.~\cite{CMS:2019ZZ} is $ 40.5 \pm 0.7 \stat \pm 1.1 \syst \pm 0.7 \lum \unit{fb}$, which agrees well with the expected value of $39.3^{+0.8}_{-0.7} \pm 0.6\unit{fb}$. 
As explained in the introduction, this fiducial cross section is valid for the current analysis.
The fiducial phase space selections are similar to the reconstruction-level selections and detailed in Table~\ref{tab:fid}. We use the notation $m_{\PZ_1,\PZ_2}$ to refer to both $m_{\PZ_1}$ and $m_{\PZ_2}$.
The MC particle-level distributions use generator-level leptons
``dressed'' by adding the momenta of generator-level photons within
$\Delta R(\ell,\PGg) < 0.1$ from the direction of the lepton.

\begin{table*}[!htb]
  \centering
  \topcaption{Particle-level selections used to define the fiducial phase space}
     \begin{tabular}{l c}
     \hline 
    Particle type & \multicolumn{1}{c}{Selection} \\
      \hline 
   &\multicolumn{1}{c}{$\ZZ$ base selection}  \\   
   \cline{2-2} \\
   Leptons & $\pt(\ell_1) > 20\GeV$ \\
           & $\pt(\ell_2) > 10\GeV$ \\
           & $\pt(\ell) > 5\GeV$ \\
           & $\abs{\eta(\ell)} < 2.5$ \\ \\
   $\PZ$ and $\ZZ$ & $40 < m_{\PZ_1} < 120\GeV$, $4 < m_{\PZ_2} < 120\GeV$ \\ 
                   & $m_{\ell\ell} > 4\GeV$ (any oppositely charged same-flavor pair) \\ \\
   Jets     & $\pt(\mathrm{j}) > 30\GeV$   \\
            & $\abs{\eta(\mathrm{j})} < 4.7$ \\
            & $\Delta R(\ell,\mathrm{j}) > 0.4$ for each $\ell, \mathrm{j}$ \\ \\ 
  \hline 
                   &\multicolumn{1}{c}{On-shell $\ZZ$ region}  \\ 
                   \cline{2-2} \\
   $\PZ$ and $\ZZ$ &\multicolumn{1}{c}{$\ZZ$ base selection + $60 < m_{\PZ_1,\PZ_2} < 120\GeV$}    \\  \\ 
  \hline 
                  &\multicolumn{1}{c}{Full $m_{4\ell}$ range}  \\ 
                  \cline{2-2} \\
  $\PZ$ and $\ZZ$ &\multicolumn{1}{c}{$\ZZ$ base selection + $m_{4\ell} > 80\GeV$}      \\

  \hline      
     \end{tabular}
        \label{tab:fid}
  \end{table*}
  
In constructing the response matrix, there are MC events that pass the reconstruction-level
selections, but do not have corresponding events at particle level that pass the fiducial
selections. In the unfolding method used,
these out-of-fiducial events are treated as background events that equivalently 
propagate from an additional particle-level bin to the reconstruction-level bins. The size of the contribution of these out-of-fiducial events can
be up to $15\%$ for events with at least one jet. In addition, the
nonprompt and VVV background events are also added to the
out-of-fiducial events.

The systematic uncertainties are propagated through the unfolding by reevaluating
the response matrix with the sample used in building the matrix shifted or reweighted to reflect a
one standard deviation variation in the quantity of interest. The resulting difference in the final normalized unfolded
distributions is taken as the uncertainty related to that quantity. 

The systematic uncertainty in the trigger efficiency is estimated to be 2\%, and cancels out in normalized differential cross sections.
To evaluate uncertainties associated with lepton efficiencies, the response matrix is reevaluated using lepton efficiency correction factors
varied up and down by the tag-and-probe~\cite{CMS:2011aa} fit uncertainties. Electrons and
muons are treated separately, and all leptons of the same type are treated as
correlated. For the uncertainties associated with the JES and jet energy resolution, the jet \pt is varied by shifting the JES and the spreading up and down by their uncertainties, and the response matrix is reevaluated.

The uncertainty in lepton fake rates is 40\%, and is dominated by the statistical uncertainty but
also includes systematic uncertainties associated with the underlying physics processes
between events in the $3\ell$ and $4\ell$ control regions.
The reducible background is varied up and down by the lepton fake rate uncertainty (40\%) and the unfolding is repeated to estimate the associated uncertainty from the difference between the normalized distributions.

The pileup uncertainty is evaluated by recomputing the response matrix with the total inelastic cross section~\cite{CMS:2018mlc}, which
defines the pileup weights applied to MC, varied up and down by
4.6\%. The uncertainty associated with the luminosity is evaluated by reevaluating the response matrix with the simulation normalized to the integrated
luminosity varied up and down by its total uncertainty, which is 1.2, 2.3 and 2.5\% for 2016, 2017 and 2018, respectively. It is small as expected due to the cancellation from the normalization by the fiducial cross section. 

The uncertainty arising from generator-specific modeling differences is evaluated from the difference
between the measurements unfolded with the response matrix based on the $\qq  \to  \ZZ$ sample simulated by \MGvATNLO and the \POWHEG sample. 

The PDF and related strong coupling ($\alpS$) uncertainties are evaluated by reweighting the \MGvATNLO sample to PDF and $\alpS$ variations, and then redoing the unfolding and combining the results according to the procedure described in Ref.~\cite{Butterworth:2015oua}. 
For the renormalization ($\mu_\mathrm{R}$) and factorization ($\mu_\mathrm{F}$) scales (QCD scales) uncertainties, the response matrix is reevaluated with the \MGvATNLO $\qq  \to  \ZZ$ sample reweighted to
reflect the distribution with $\mu_\mathrm{F}$ and $\mu_\mathrm{R}$ independently varied up
and down by a factor of 2. All combinations are considered except those in which $\mu_\mathrm{F}$ and $\mu_\mathrm{R}$ differ by a factor of four, 
and the envelope of all variations is used.

The normalization of the \MCFM sample ($\Pg\Pg \! \to \! \ZZ$) is varied by the scale and
PDF uncertainties of its cross section (${}^{+18\%}_{-14\%}$), and the
resulting difference between the normalized distributions is used.

The contributions of each source of systematic uncertainty to the final results are 
summarized in Tables~\ref{table:systematics1}--\ref{table:systematics3}. The numbers in these tables are only indicative. They are estimated by varying each source and obtaining the difference in the normalized unfolded distributions. Each number in the tables is not showing an estimate of uncertainty per bin, but an estimate of the portion of the uncertainty contribution per distribution, given by 
\begin{equation}
\frac{\sum\limits_{i=1}^{N_{\text{bins}}} \abs{h_{\text{central}}(i)-h_{\text{varied}}(i)}}{\sum\limits_{i=1}^{N_{\text{bins}}} h_{\text{central}}(i)}
\end{equation}
where $h_{\text{central}}$ and $h_{\text{varied}}$ are the 
central and varied histograms, respectively, both with sum of bin contents normalized to 1, and $N_{\text{bins}}$ is the total number of bins. There are, in general, two estimates from up/down variations and the larger one is used.

\begin{table}[htb]
\centering
\topcaption{
  Contributions of each source of systematic uncertainty to the
  normalized differential cross section measurements of jet variables. Uncertainties depend on the distributions and are listed as a range.  
}
\begin{tabular}{lc}
\hline
Systematic source &  Uncertainty range  \\
\hline                                                                           
Electron efficiency                          & 0.13--0.30\% \\ 
Muon efficiency                              & 0.02--0.08\% \\ 
Jet energy resolution                        & 1.65--3.85\% \\ 
JES correction                               & 0.93--5.32\% \\ 
Reducible background                         & 0.05--0.43\% \\ 
Pileup                                       & 0.04--1.08\% \\ 
Luminosity                                   & $<0.03\%$    \\ 
$\qq  \to  \ZZ$ MC choice                    & 0.52--4.52\% \\ 
$\Pg\Pg  \to  \ZZ$ cross section             & 0.01--0.19\% \\
QCD scales                                   & 0.16--0.82\% \\
PDF                                          & 0.05--0.12\% \\ 
PDF $\alpS$                                  & 0.01--0.10\% \\ 
\hline

\end{tabular}
\label{table:systematics1}
\end{table}

\begin{table}[htb]
\centering
\topcaption{
  The contributions of each source of systematic uncertainty in the
  normalized differential cross sections measurements as a function of $m_{4\ell}$ with jet multiplicity from 0 to 3 and more, in events satisfying $60 < m_{\PZ_1, \PZ_2} < 120\GeV$. 
  }

\begin{tabular}{lcccccc}  
\hline
Systematic source &  $m_{4\ell}$ with all jets & 0 jet & 1 jet & 2 jets & 3 and more jets \\
\hline                                                                           
Electron efficiency                          & 0.42\% & 0.38\% & 0.66\% & 0.36\% & 0.26\% \\ 
Muon efficiency                              & 0.05\% & 0.06\% & 0.07\% & 0.09\% & 0.08\% \\ 
Jet energy resolution                        &  \NA   & 0.07\% & 1.72\% & 1.65\% & 0.80\% \\ 
JES correction                               &  \NA   & 0.17\% & 1.77\% & 1.95\% & 0.97\% \\ 
Reducible background                         & 0.18\% & 0.18\% & 0.32\% & 0.33\% & 0.96\% \\ 
Pileup                                       & 0.02\% & 0.05\% & 0.11\% & 0.13\% & 0.35\% \\ 
Luminosity                                   & 0.01\% & 0.01\% & 0.02\% & 0.02\% & 0.05\% \\ 
$\qq  \to  \ZZ$ MC choice                    & 0.35\% & 0.65\% & 0.94\% & 0.48\% & 0.35\% \\ 
$\Pg\Pg  \to  \ZZ$ cross section             & 0.02\% & 0.03\% & 0.09\% & 0.06\% & 0.09\% \\ 
QCD scales                                   & 0.15\% & 0.16\% & 0.58\% & 0.54\% & 0.62\% \\ 
PDF                                          & 0.05\% & 0.05\% & 0.15\% & 0.15\% & 0.21\% \\ 
PDF $\alpS$                                  & 0.02\% & 0.01\% & 0.05\% & 0.03\% & 0.02\% \\
\hline

\end{tabular}
\label{table:systematics2}
\end{table}

\begin{table}[htb]
\centering
\topcaption{
  The contributions of each source of systematic uncertainty in the
  normalized differential cross sections measurements as a function of $m_{4\ell}$ with jet multiplicity from 0 to 3 and more, in events from the full $m_{4\ell}$ range. 
  }

\begin{tabular}{lcccccc}
\hline
Systematic source &  $m_{4\ell}$ with all jets & 0 jet & 1 jet & 2 jets & 3 and more jets  \\
\hline                                                                           
Electron efficiency                          & 2.12\% & 2.55\% & 2.28\% & 1.77\% & 1.46\% \\ 
Muon efficiency                              & 0.71\% & 0.78\% & 0.92\% & 0.79\% & 0.42\% \\ 
Jet energy resolution                        &  \NA   & 0.11\% & 1.73\% & 2.63\% & 2.32\% \\ 
JES correction                               &  \NA   & 0.33\% & 1.64\% & 3.01\% & 2.02\% \\ 
Reducible background                         & 2.22\% & 2.19\% & 2.88\% & 3.40\% & 5.09\% \\ 
Pileup                                       & 0.21\% & 0.28\% & 0.19\% & 0.32\% & 0.52\% \\ 
Luminosity                                   & 0.12\% & 0.12\% & 0.16\% & 0.17\% & 0.25\% \\ 
$\qq  \to  \ZZ$ MC choice                    & 0.57\% & 0.48\% & 1.22\% & 3.07\% & 4.21\% \\ 
$\Pg\Pg  \to  \ZZ$ cross section             & 0.10\% & 0.18\% & 0.61\% & 0.80\% & 0.46\% \\ 
QCD scales                                   & 0.27\% & 0.25\% & 0.67\% & 1.25\% & 1.86\% \\ 
PDF                                          & 0.07\% & 0.09\% & 0.20\% & 0.23\% & 0.28\% \\ 
PDF $\alpS$                                  & 0.08\% & 0.08\% & 0.15\% & 0.20\% & 0.28\% \\
\hline

\end{tabular}
\label{table:systematics3}
\end{table}

\clearpage

\section{Results}

\subsection{Differential distributions}

\begin{figure}[htbp]
\centering
\includegraphics[width=0.45\textwidth]{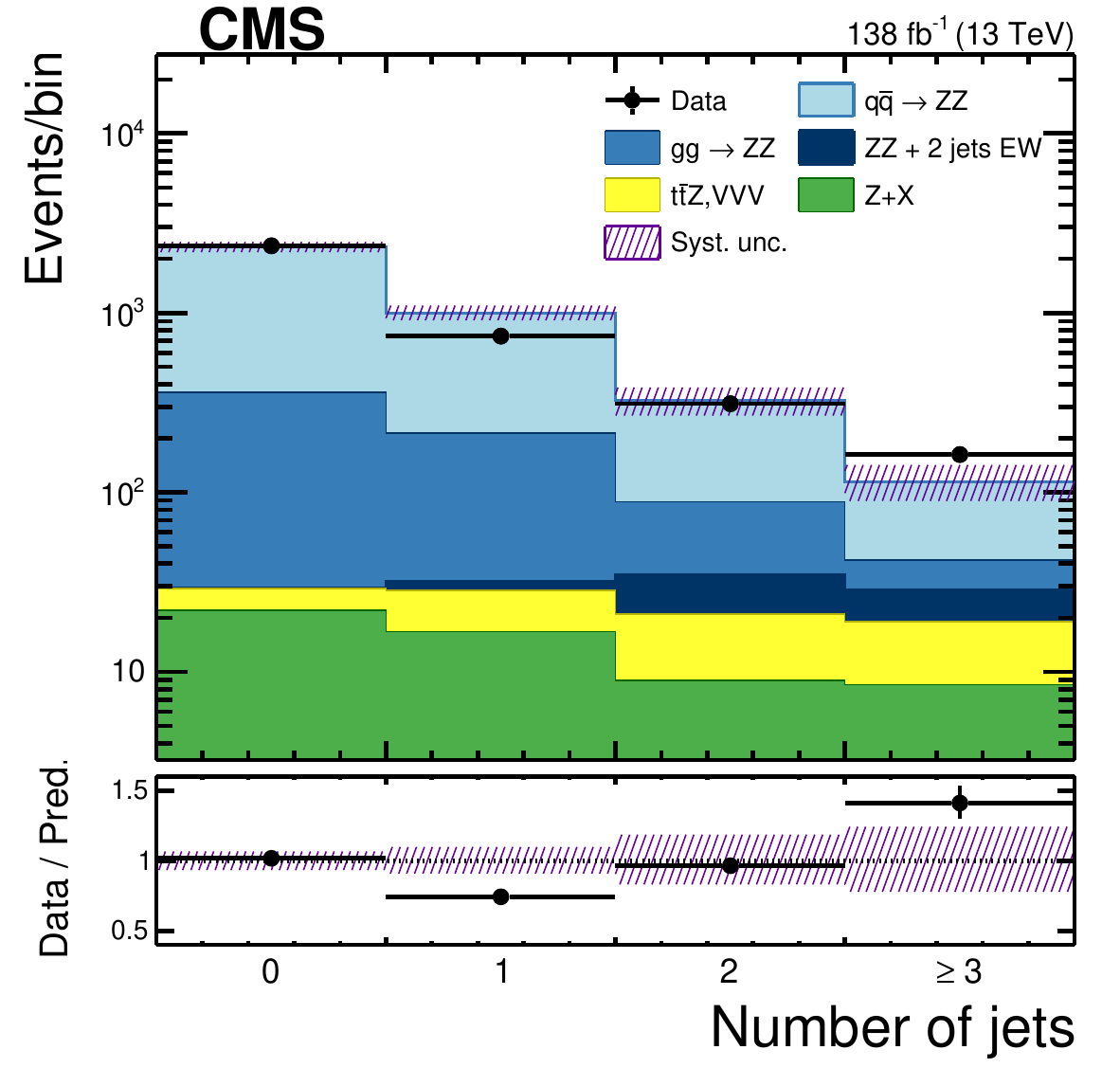}
\includegraphics[width=0.45\textwidth]{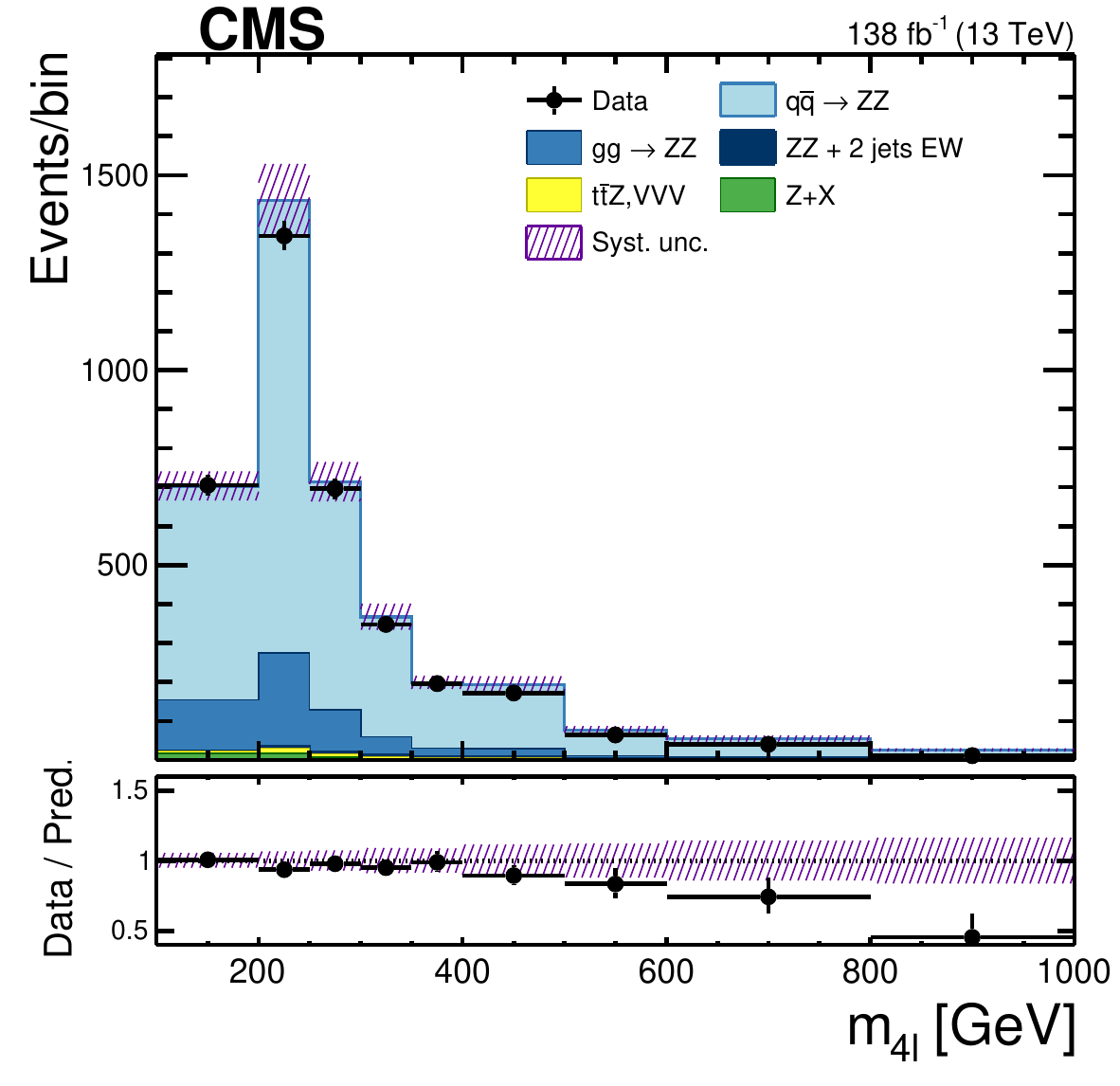}
\caption{
Distribution of the number of jets with $\pt>30\GeV$ (left) and of $m_{\ZZ}$ (right)
for $\ZZ$+jets events with $60 < m_{\PZ_1, \PZ_2} < 120\GeV$ for the combined 4$\Pe$, 4$\PGm$, and 2$\Pe$2$\PGm$ decay channels. 
Points represent the data, vertical bars the statistical uncertainties, and shaded histograms represent
the expected standard model predictions and reducible background estimated from
data. The purple band of slashes represents the systematic uncertainties in the predictions, which includes systematic uncertainties associated with trigger efficiency, lepton efficiencies, jet energy correction and jet energy resolution, pileup, luminosity, Monte Carlo generator choice, $\Pg\Pg \! \to \! \ZZ$ cross section, and reducible background. The overflow is included in the last bin of the distributions. 
\label{fig:RECO1}
}
\end{figure}

\begin{figure}[htbp]
\centering
\includegraphics[width=0.45\textwidth]{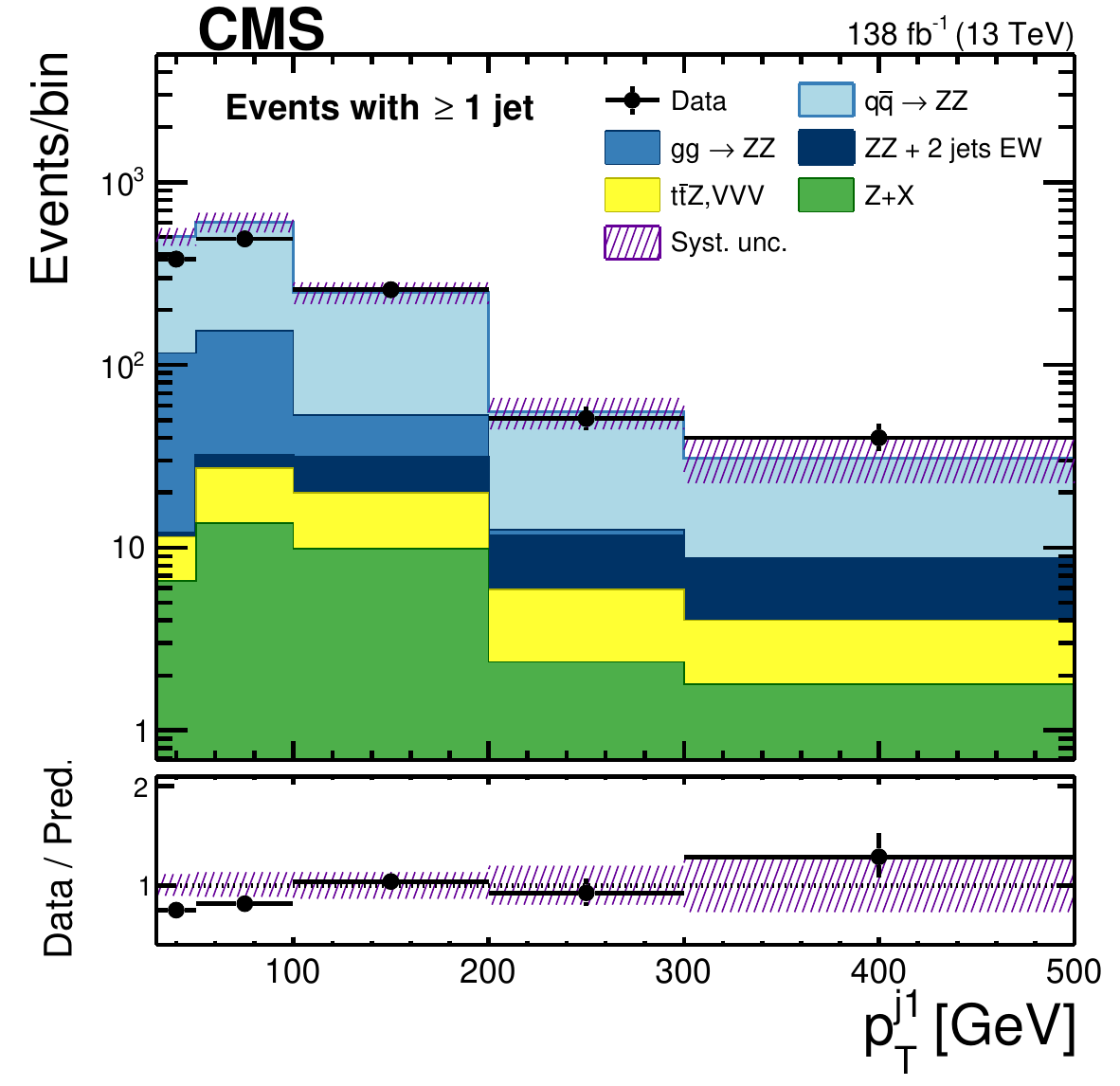}
\includegraphics[width=0.45\textwidth]{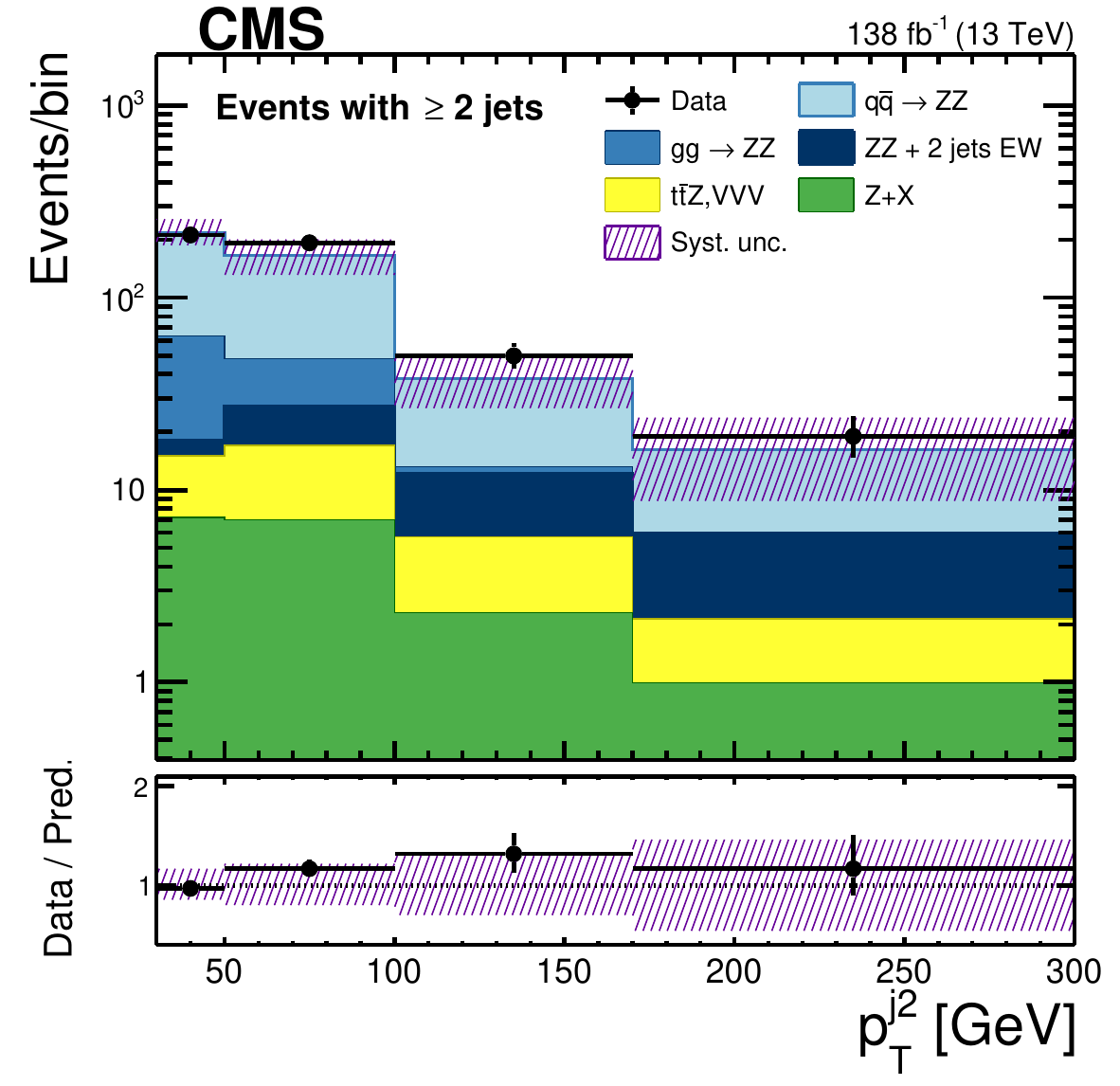}
\includegraphics[width=0.45\textwidth]{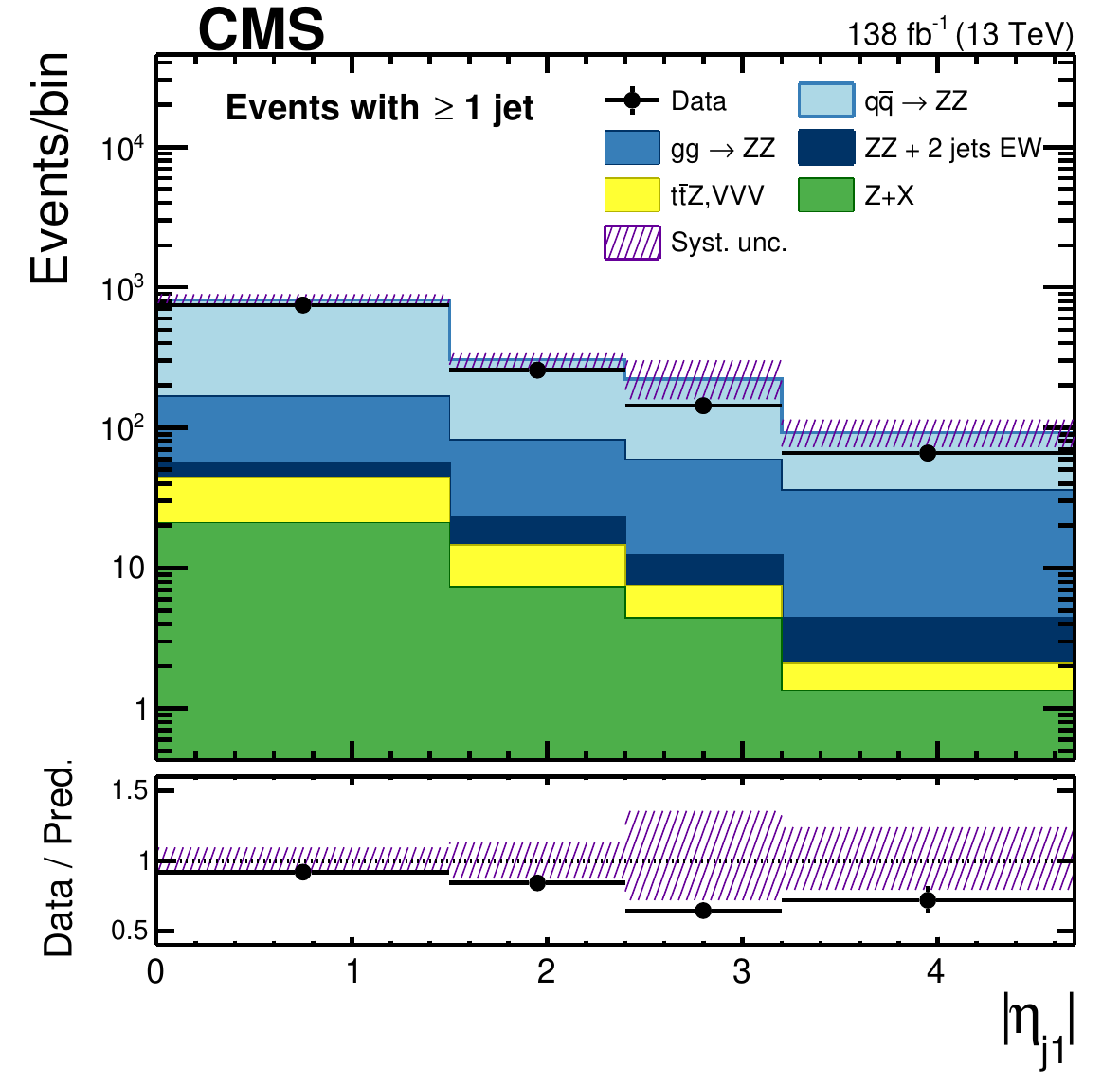}
\includegraphics[width=0.45\textwidth]{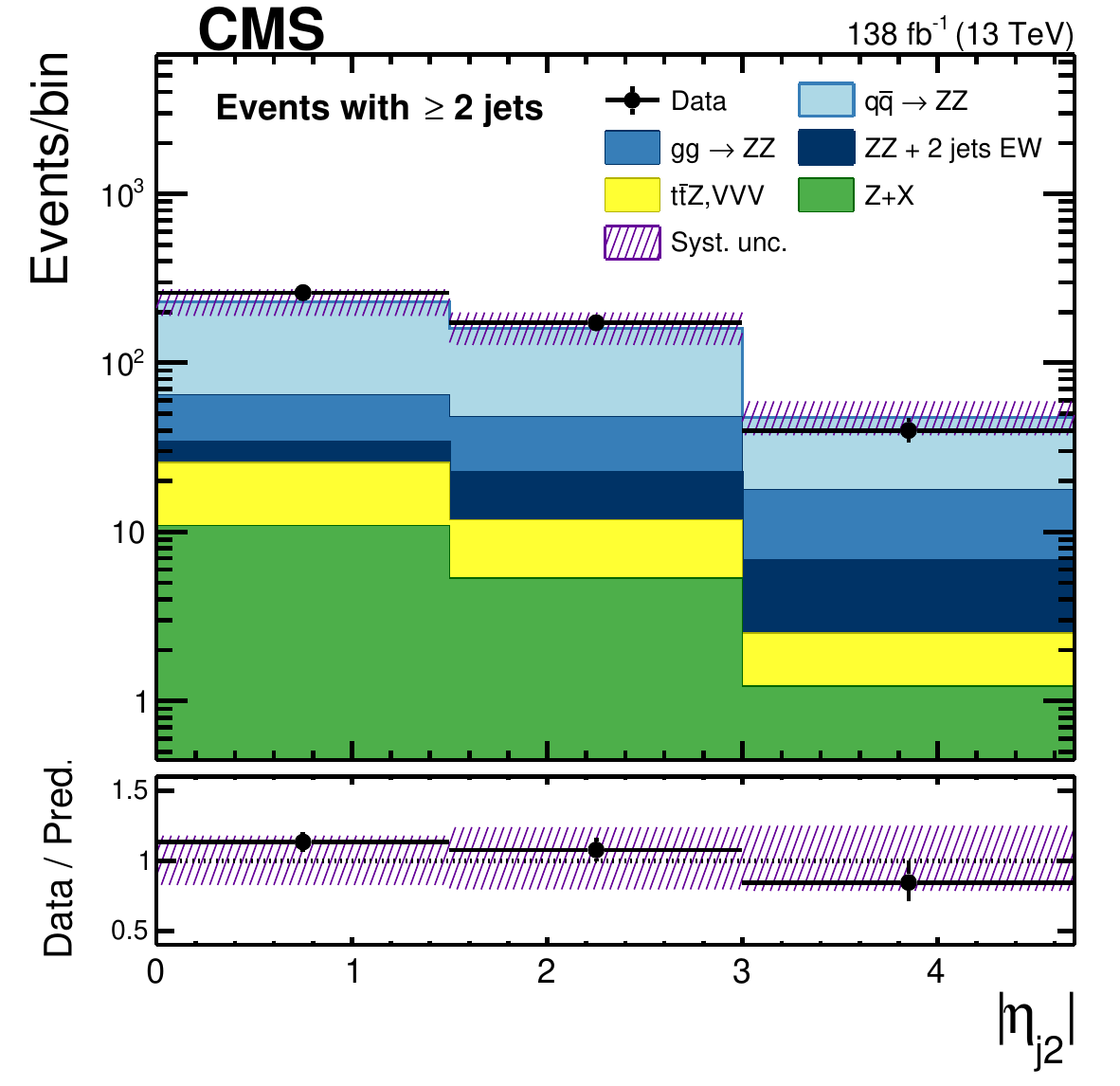}
\caption{
Distribution of the \pt of the highest-\pt jet (upper left) in events with at least one jet, and of the \pt
of the second-highest-\pt jet (upper right) in events containing at least two jets. 
The $\abs{\eta}$ distribution of the highest-\pt (lower left) and second-highest-\pt (lower right) jets. Events are required to have $60 < m_{\PZ_1, \PZ_2} < 120\GeV$.
Other details are as in the caption of Fig.~\ref{fig:RECO1}.
\label{fig:RECO2}
}
\end{figure}

Differential distributions for various reconstructed quantities are presented in this subsection. We proceed with unfolding the data to compare directly with particle-level theoretical predictions, and the results are presented in the next subsection. Figure~\ref{fig:RECO1} (left) shows the number of reconstructed jets with $\pt > 30\GeV$ for the ZZ+jets events with $60 < m_{\PZ_1, \PZ_2} < 120\GeV$. The last bin
includes all events with three
or more jets.
The description of events with three and more jets requires NNLO and even higher corrections, but there
are not enough hard jets from the matrix element in the MC samples used, therefore
the difference between data and predictions at high jets multiplicity is expected. 
The 0 and 2 jet bins are well described by the 
predictions, whereas in the 1 jet bin the predictions significantly overestimate the measured event yield. 
The $m_{4\ell}$ distribution is shown in Fig.~\ref{fig:RECO1} (right),
inclusive in the number of jets. This distribution is well described by the predictions, except for the increasing discrepancy between data
and MC towards high masses; this can be mitigated by adding the EW corrections,
as demonstrated in Ref.~\cite{Buonocore:2021fnj} and in the next subsection.

Figure~\ref{fig:RECO2} shows the \pt and $\abs{\eta}$ distributions for the highest- and second-highest-\pt jet  
in events with at least one and two jets, respectively. As expected from the distribution of the number
of jets, the predictions overestimate the measurements in the highest-\pt jet distributions. The largest difference is observed for highest-\pt jets with $\pt < 100\GeV$, whereas the
second-highest-\pt distributions are better described. Apart from the difference in the yield, the \pt distributions of both the highest-\pt and
second-highest-\pt jets show similar differential behavior with respect to the predictions (similar trend up to 300\GeV), which is demonstrated 
in the lower panels of the figures, 
where data-to-prediction ratios are presented. Similar conclusions are valid also for the $\abs{\eta}$ distributions.

\begin{figure}[htbp]
\centering
\includegraphics[width=0.45\textwidth]{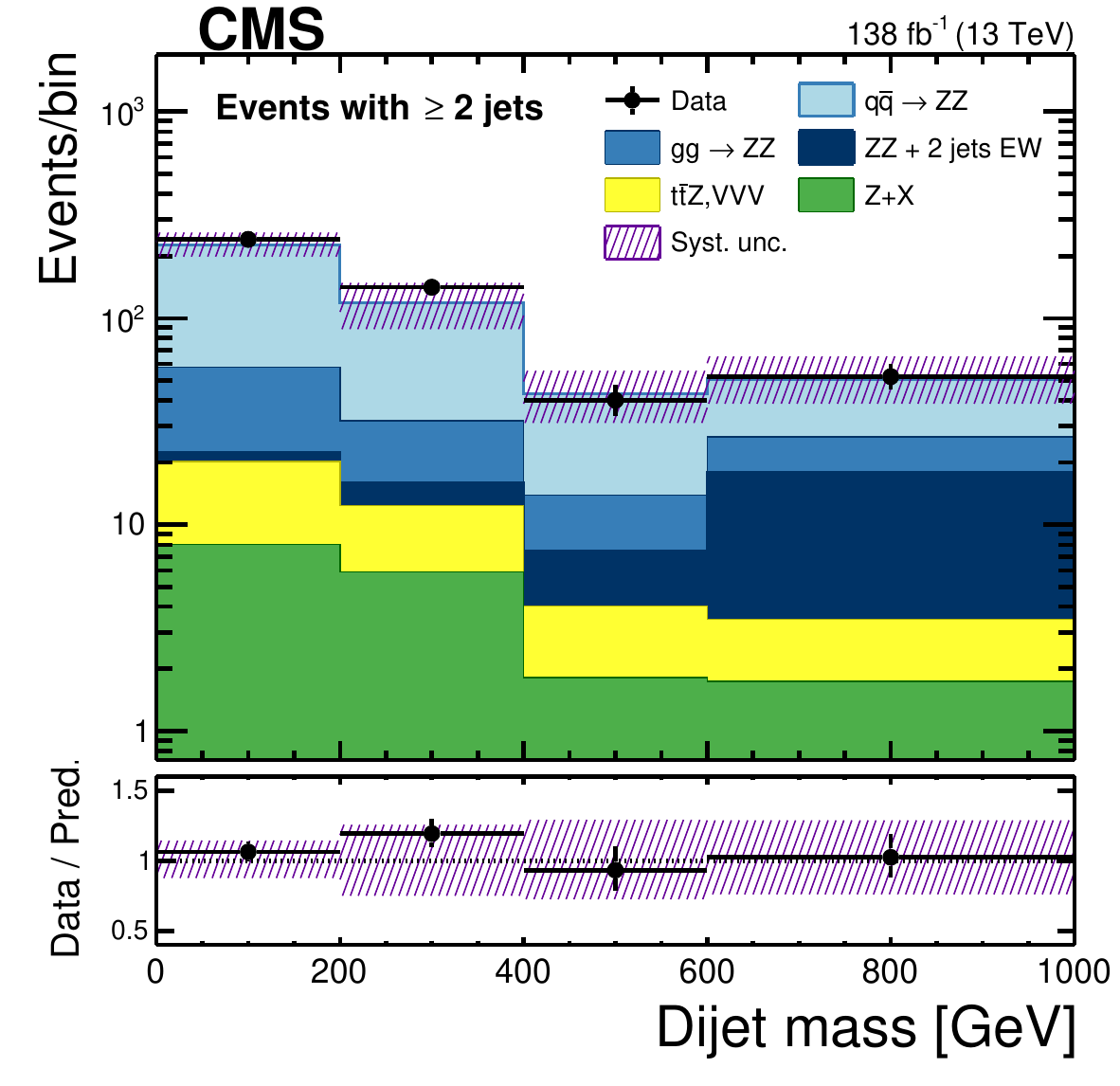}
\includegraphics[width=0.45\textwidth]{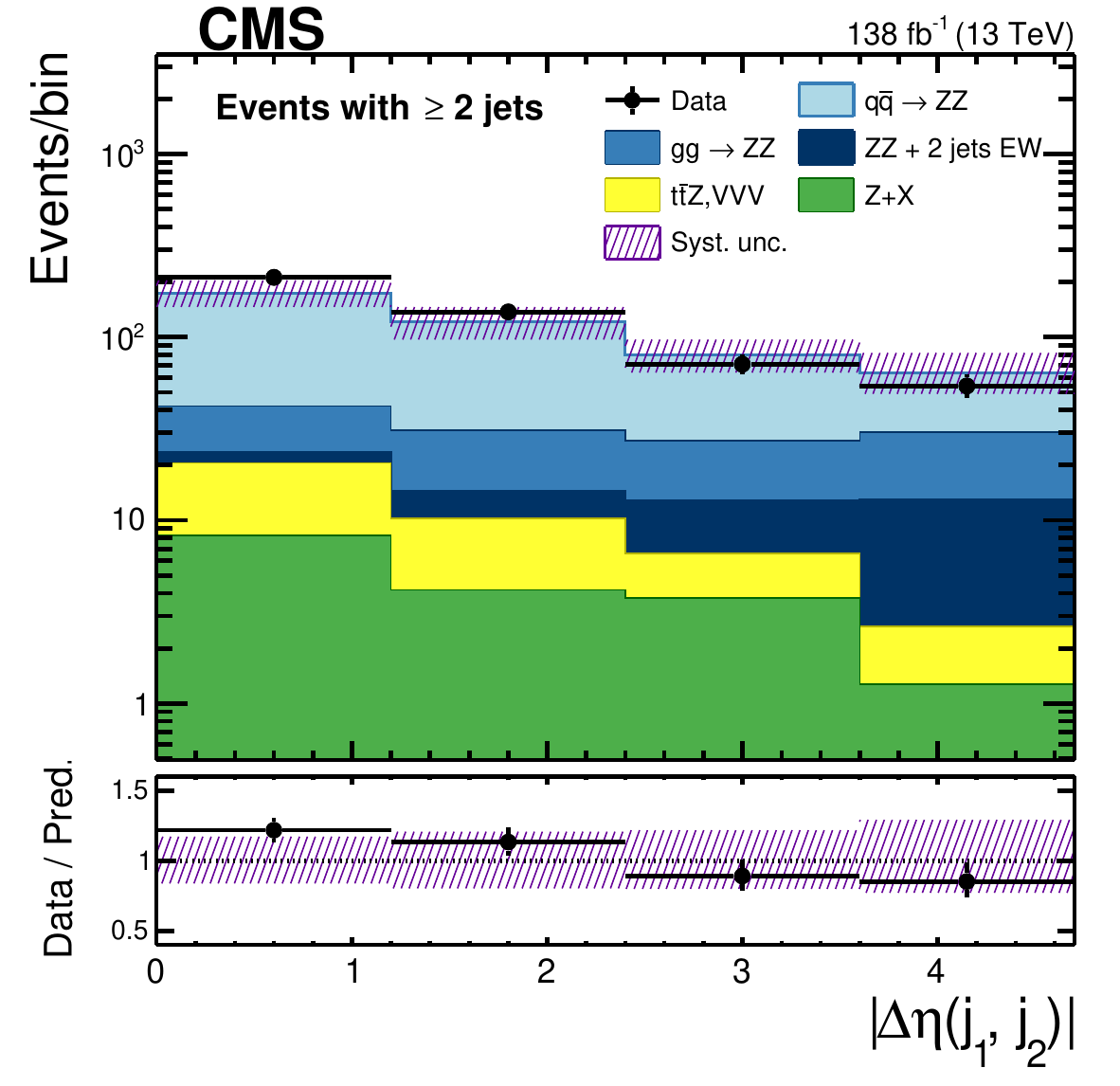}
\caption{
The dijet mass (left) and $\abs{\Delta \eta}$ (right) between the two highest-\pt jets in events with at least two jets. Events are required to have $60 < m_{\PZ_1, \PZ_2} < 120\GeV$.
Other details are as in the caption of Fig.~\ref{fig:RECO1}.
\label{fig:RECO3}
}
\end{figure}

The invariant mass of the dijet system and $\Delta \eta$ between two jets with highest \pt are among the most important dijet distributions.
The dijet mass distribution is well described by predictions, as shown in Fig.~\ref{fig:RECO3} (left), whereas
in the $\abs{\Delta \eta}$ distribution there is a small trend between data and predictions 
that can be seen in the lower panel 
of Fig.~\ref{fig:RECO3} (right).
As expected, the contribution of the EW ZZ production is increasing towards the high jet separation
and dijet mass, but still remains a small part of the total ZZ cross section.

\begin{figure}[htbp]
\centering
\includegraphics[width=0.45\textwidth]{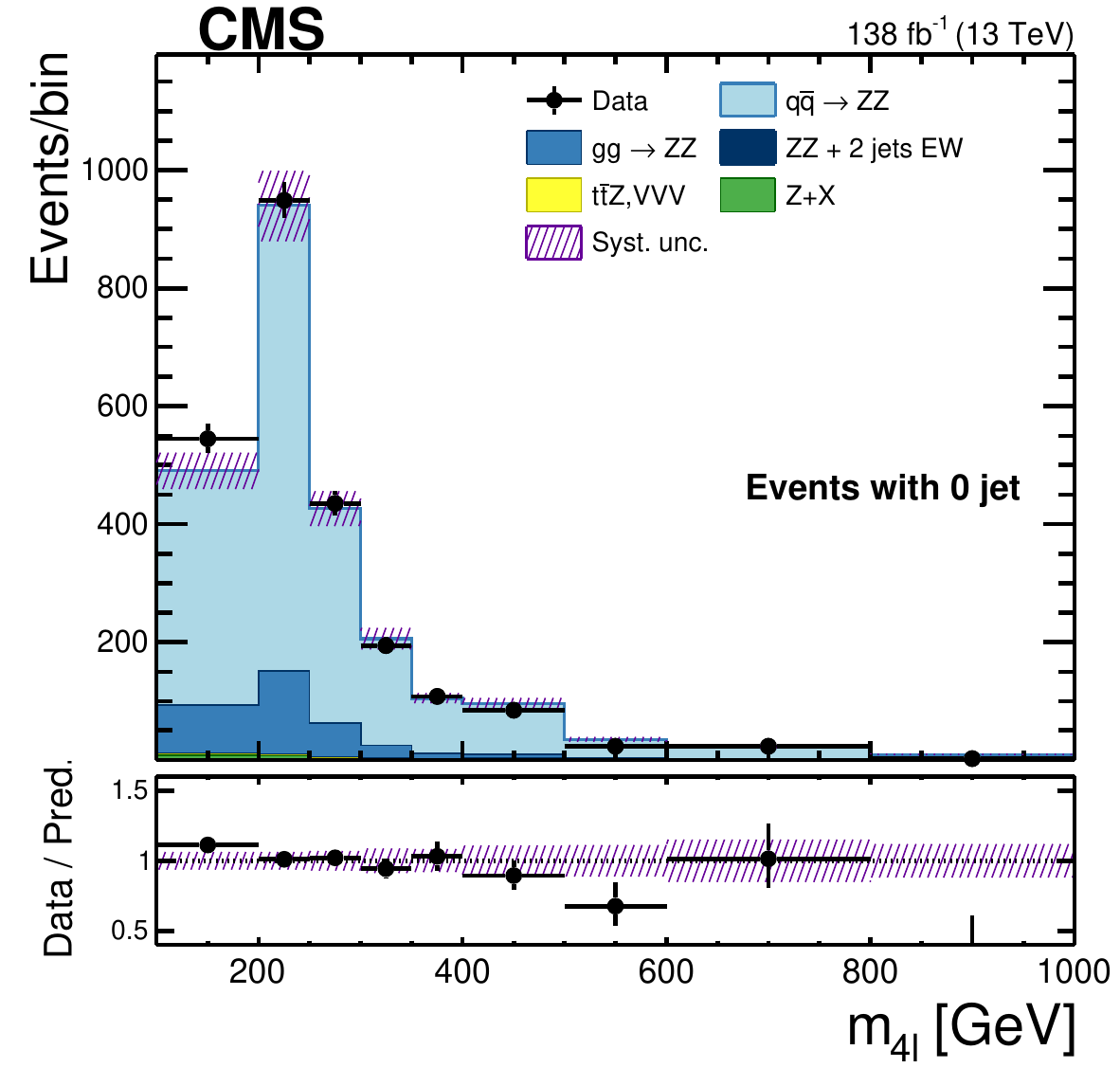}
\includegraphics[width=0.45\textwidth]{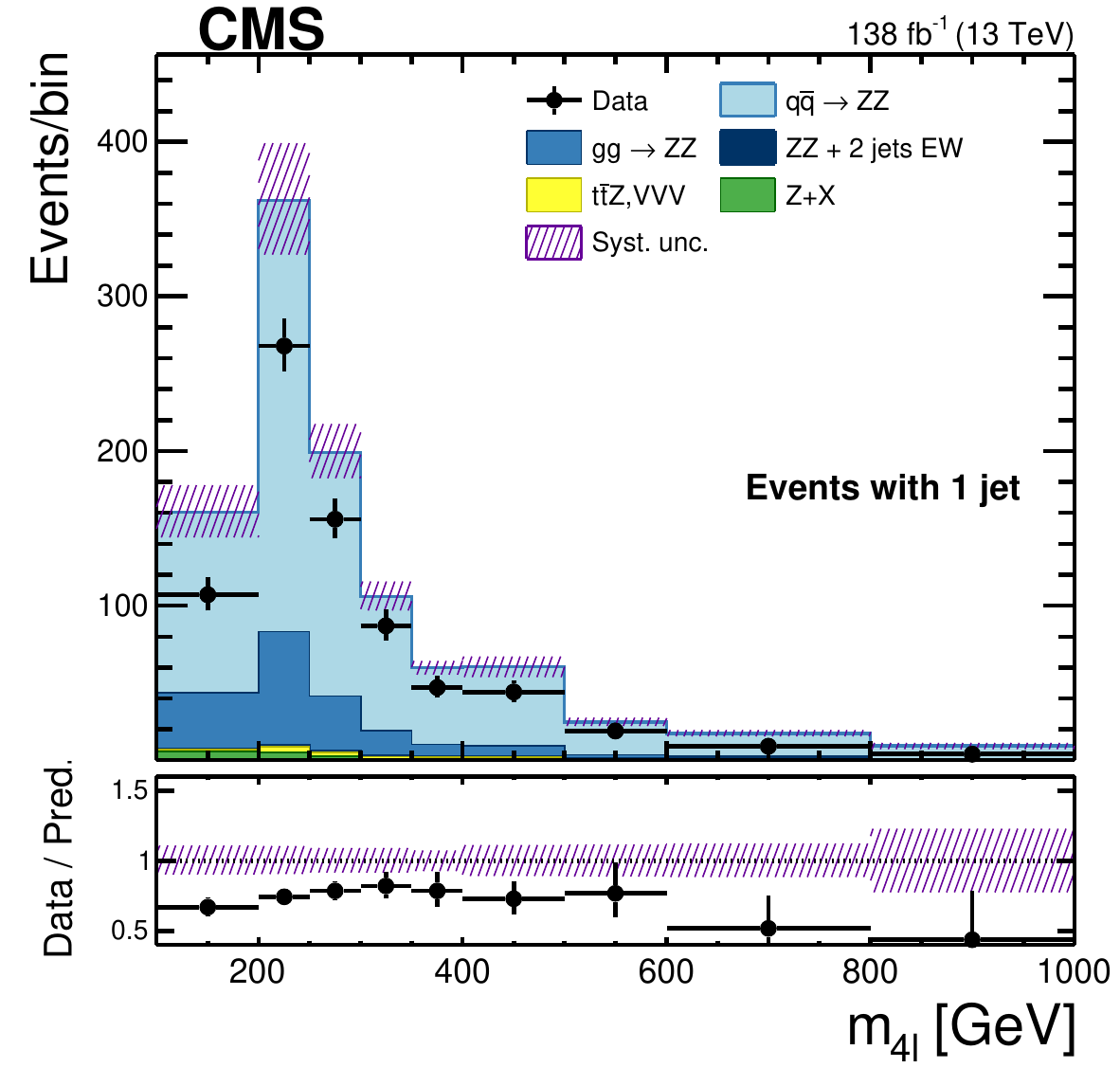}
\includegraphics[width=0.45\textwidth]{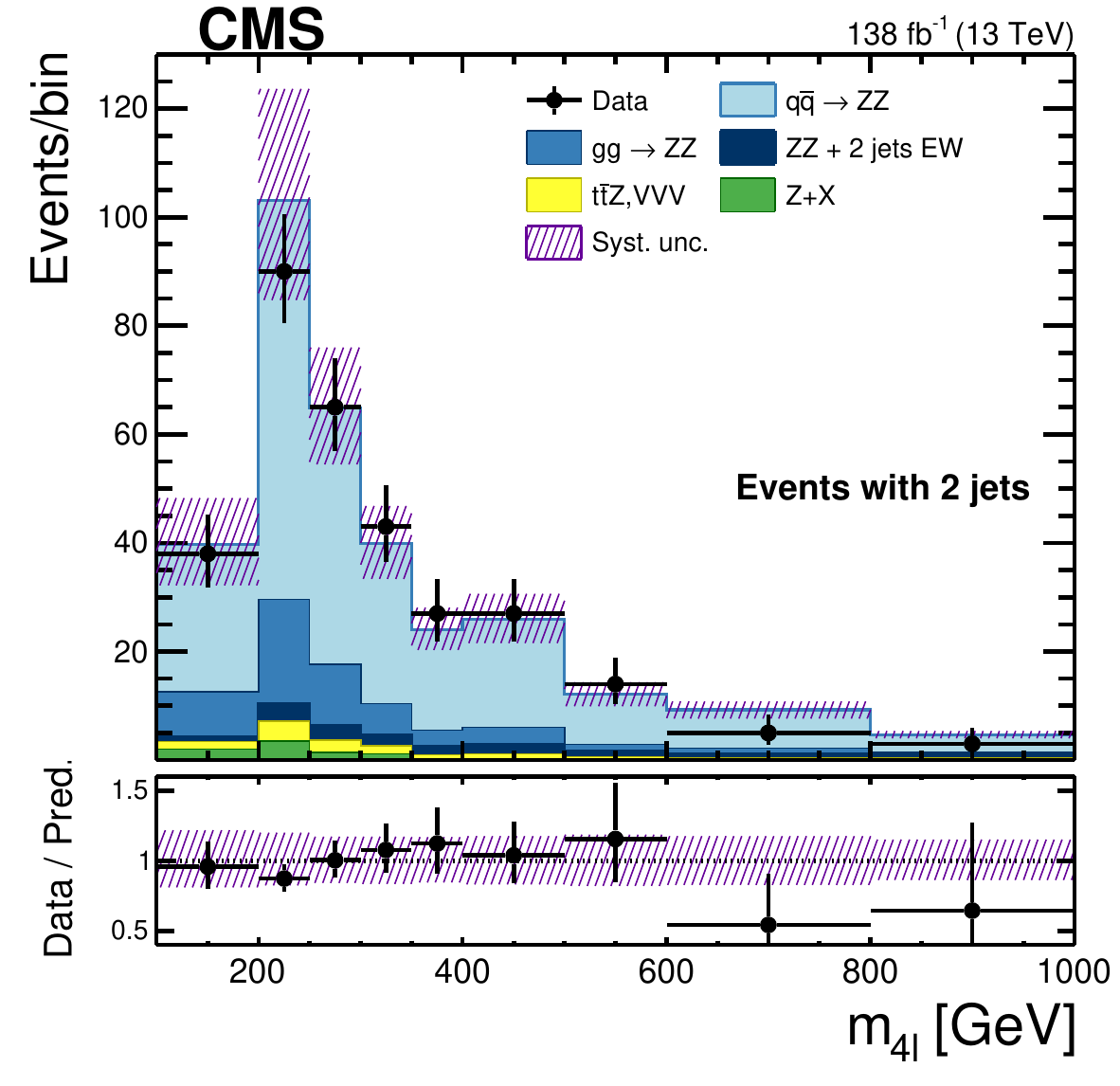}
\includegraphics[width=0.45\textwidth]{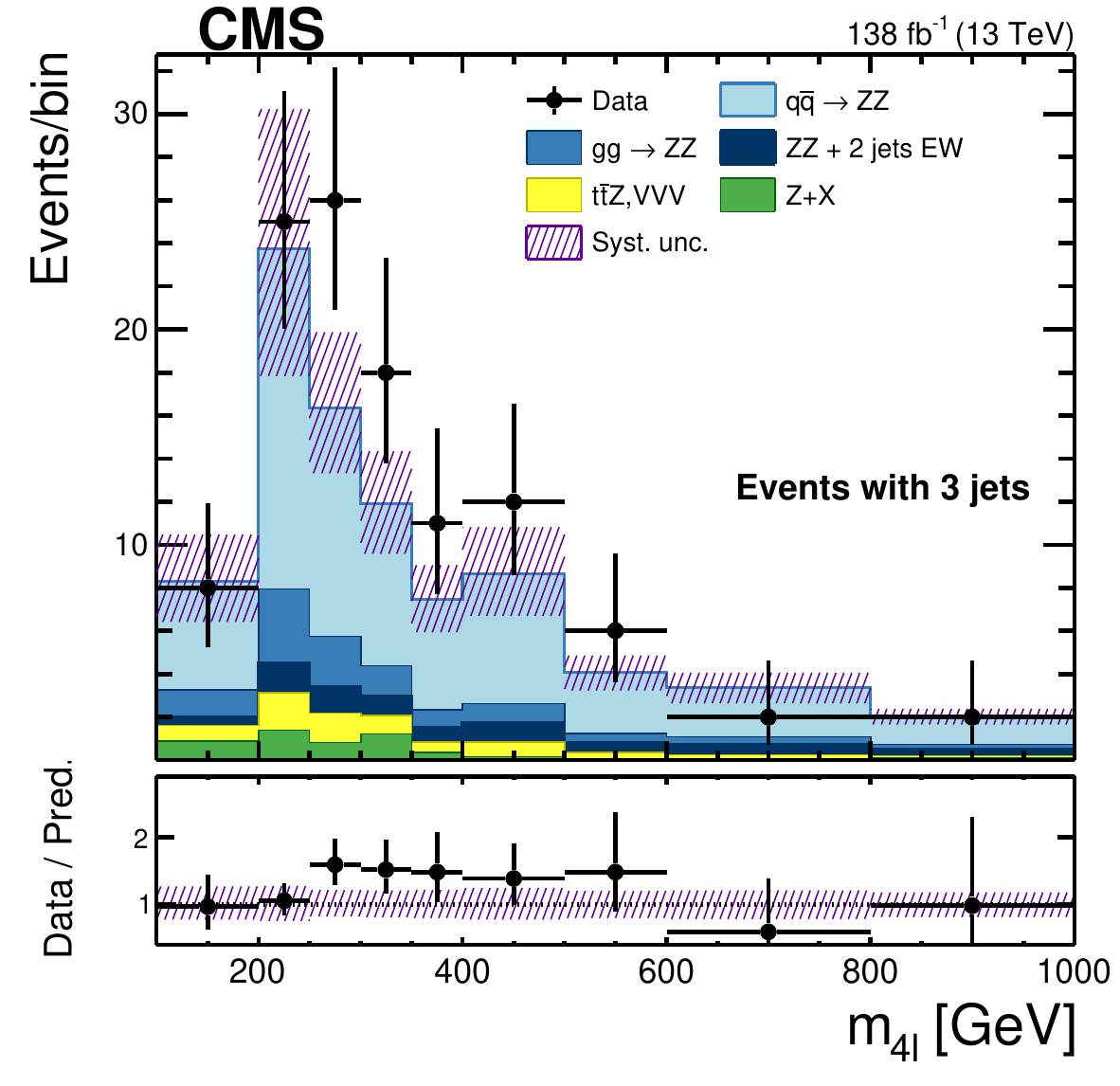}
\includegraphics[width=0.45\textwidth]{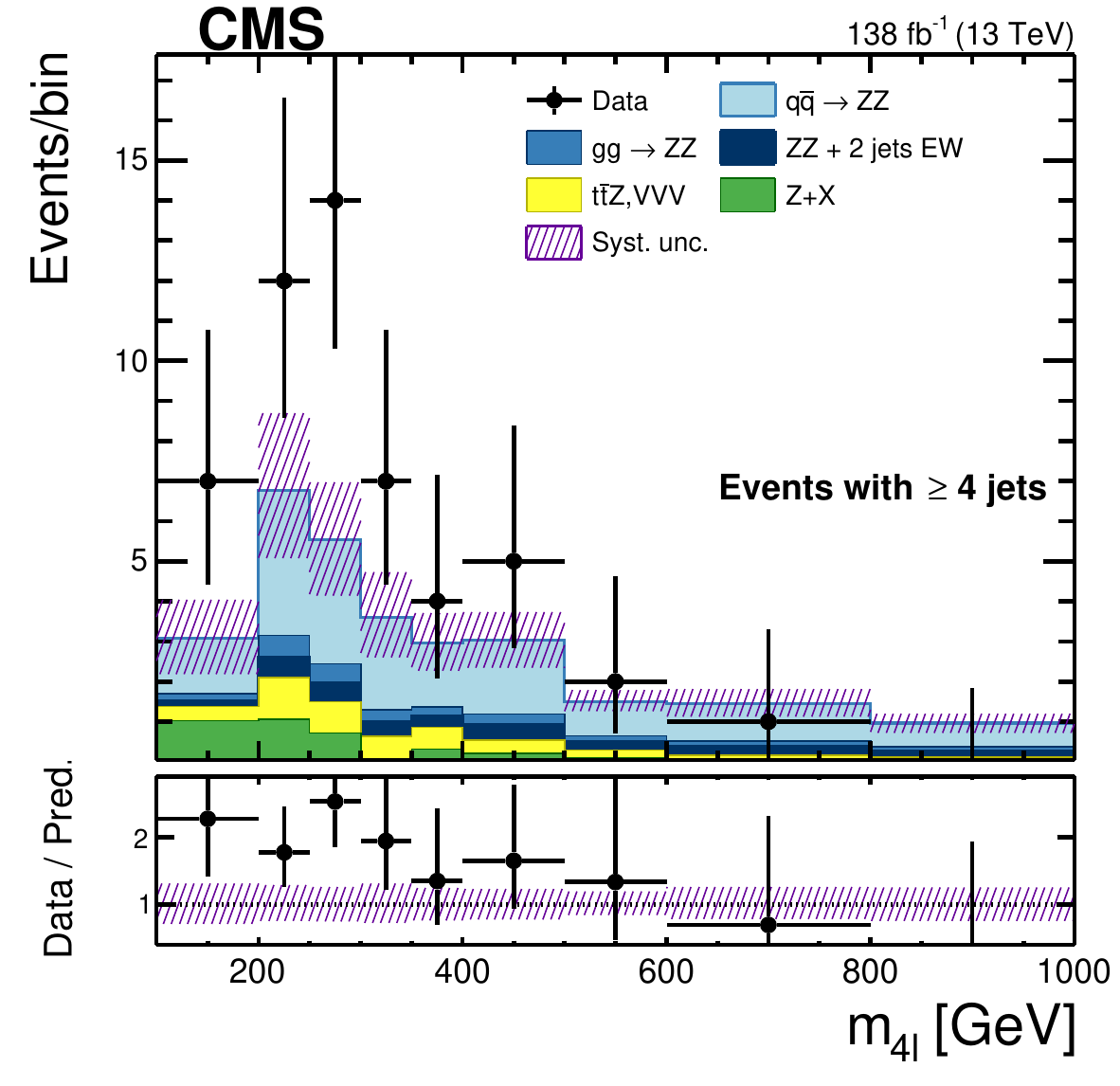}
\caption{
The $m_{4\ell}$ distributions for events with $60 < m_{\PZ_1, \PZ_2} < 120\GeV$ and different jet multiplicities.
Other details are as in the caption of Fig.~\ref{fig:RECO1}.
\label{fig:RECO4}
}
\end{figure}

\begin{figure}[htbp]
\centering
\includegraphics[width=0.45\textwidth]{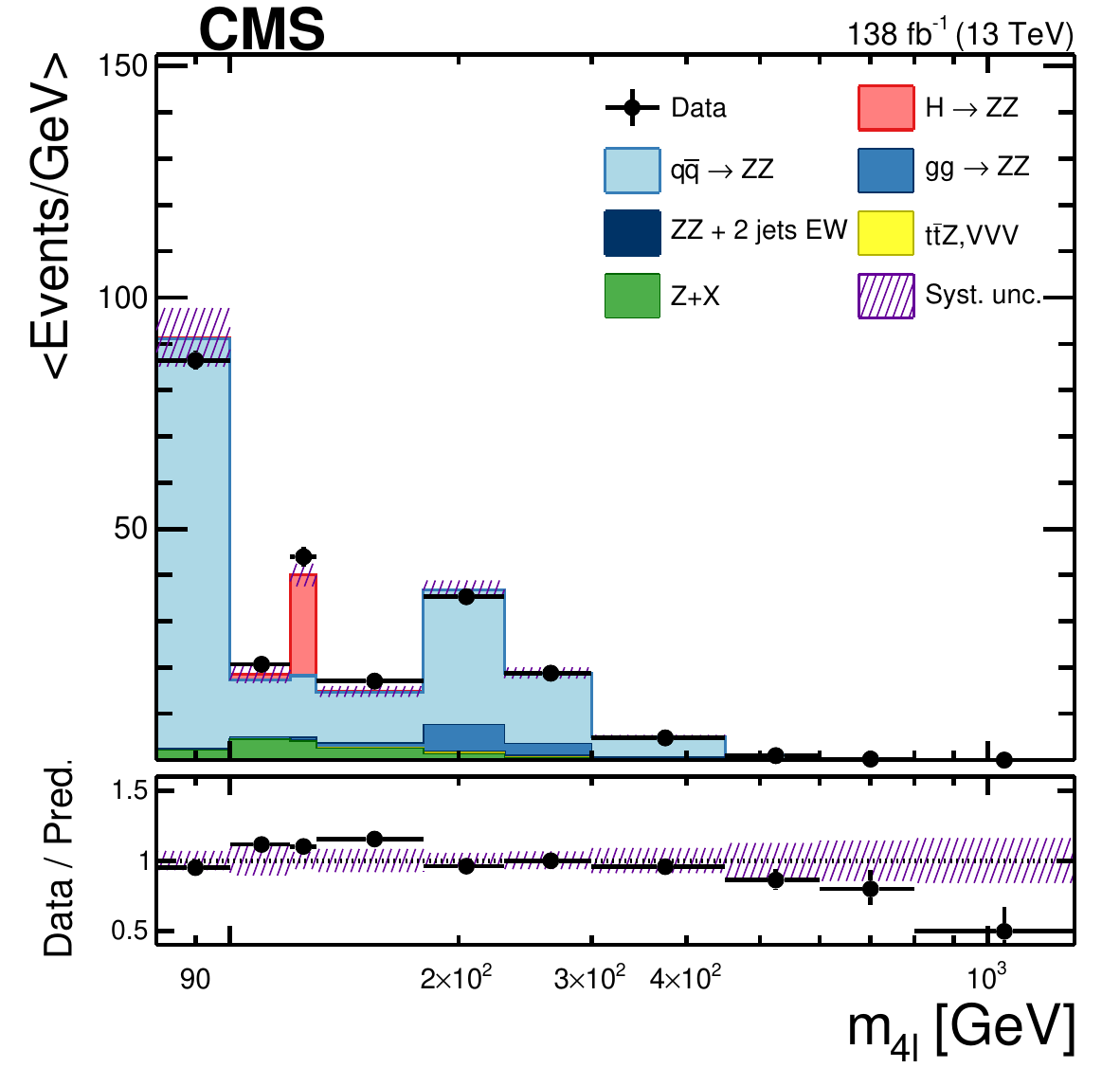}
\includegraphics[width=0.45\textwidth]{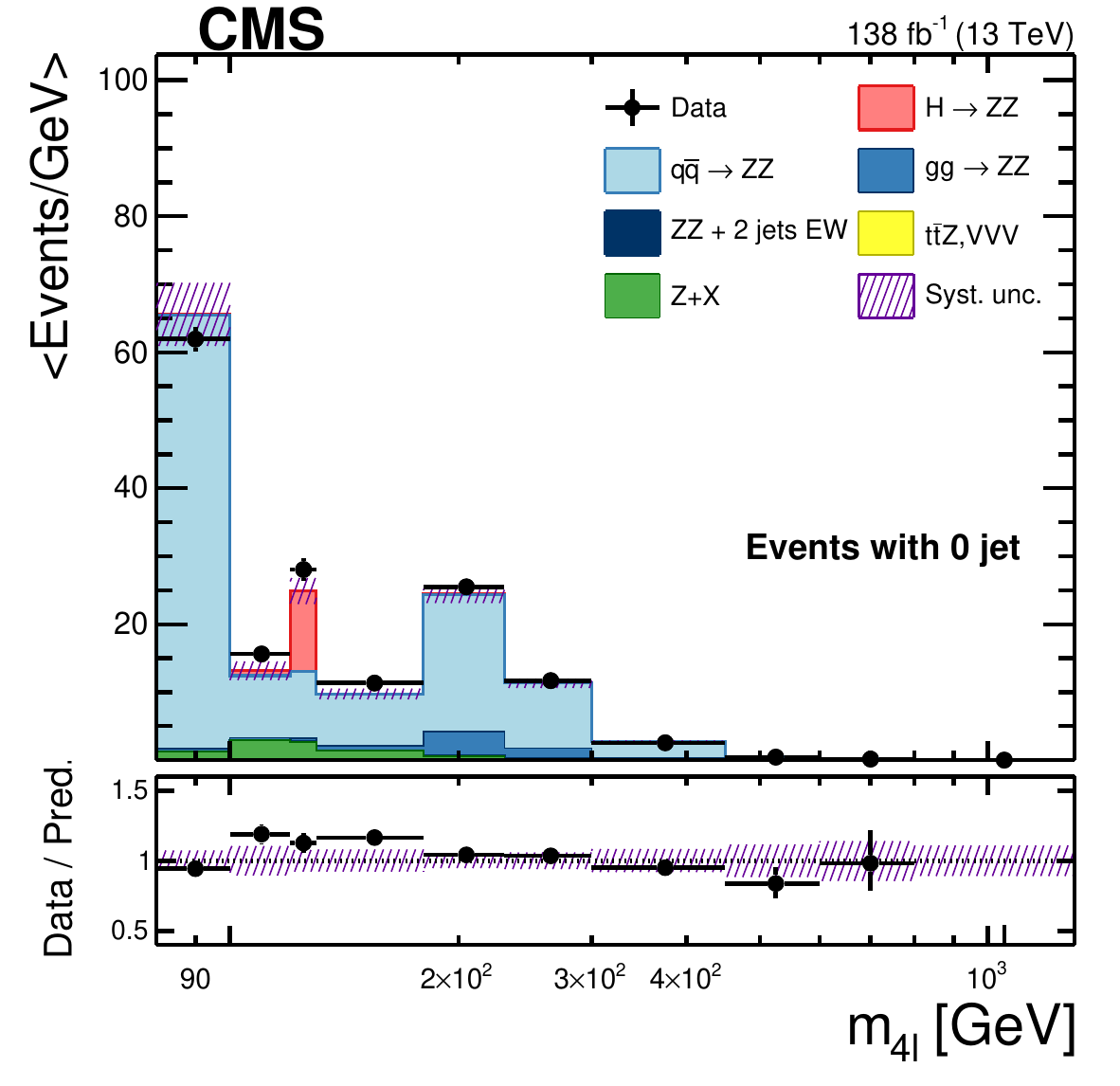}
\includegraphics[width=0.45\textwidth]{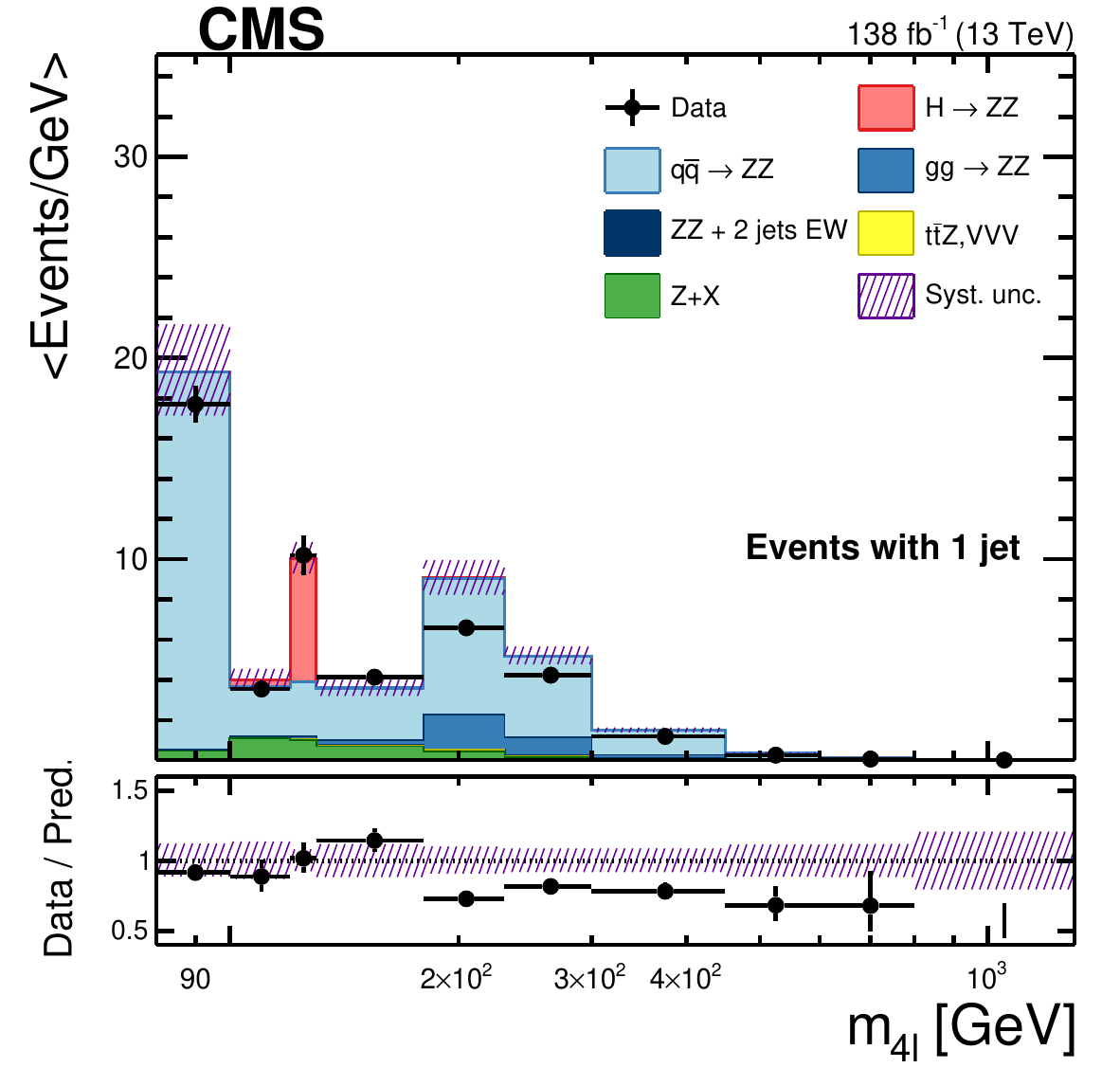}
\includegraphics[width=0.45\textwidth]{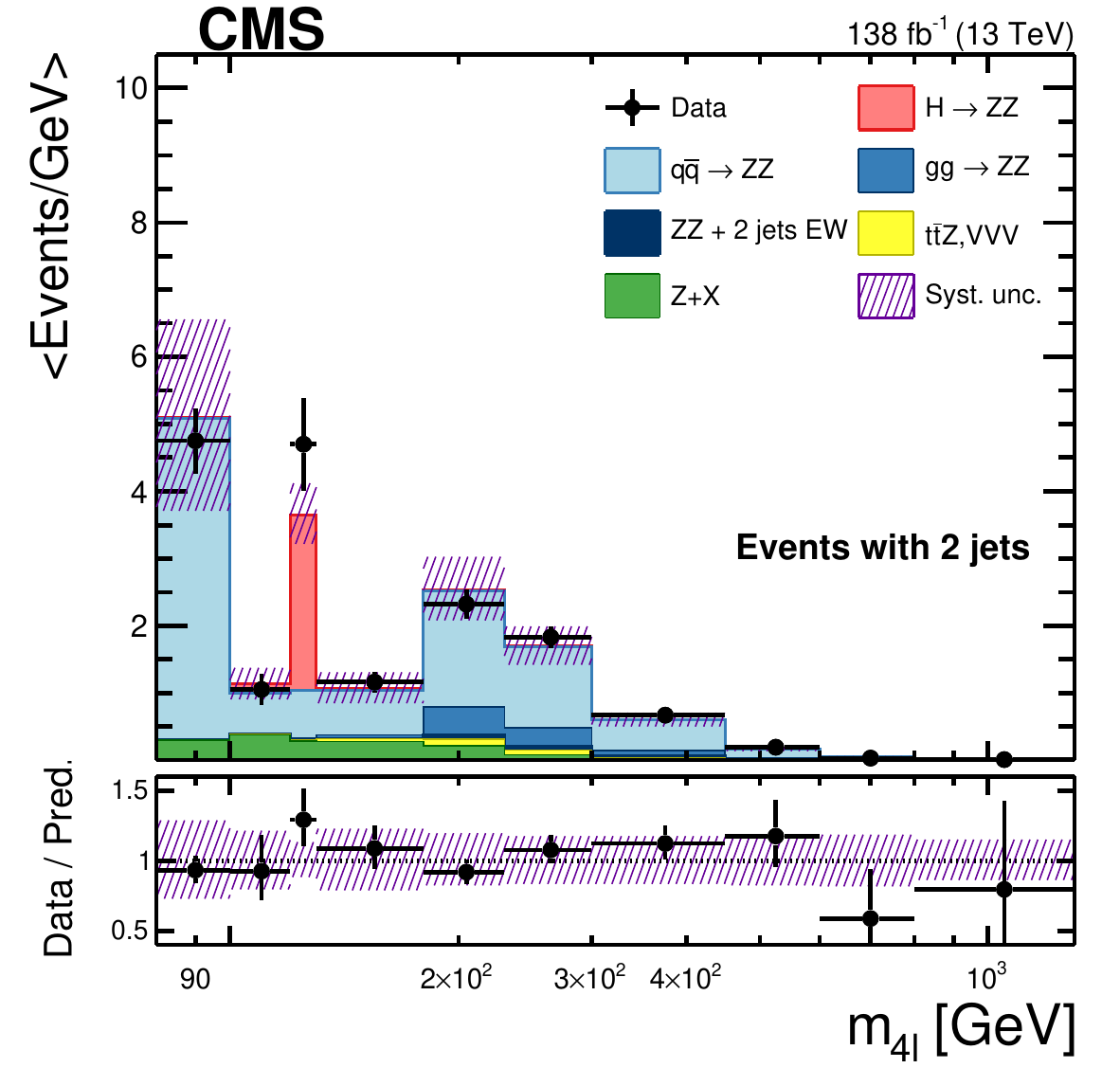}
\includegraphics[width=0.45\textwidth]{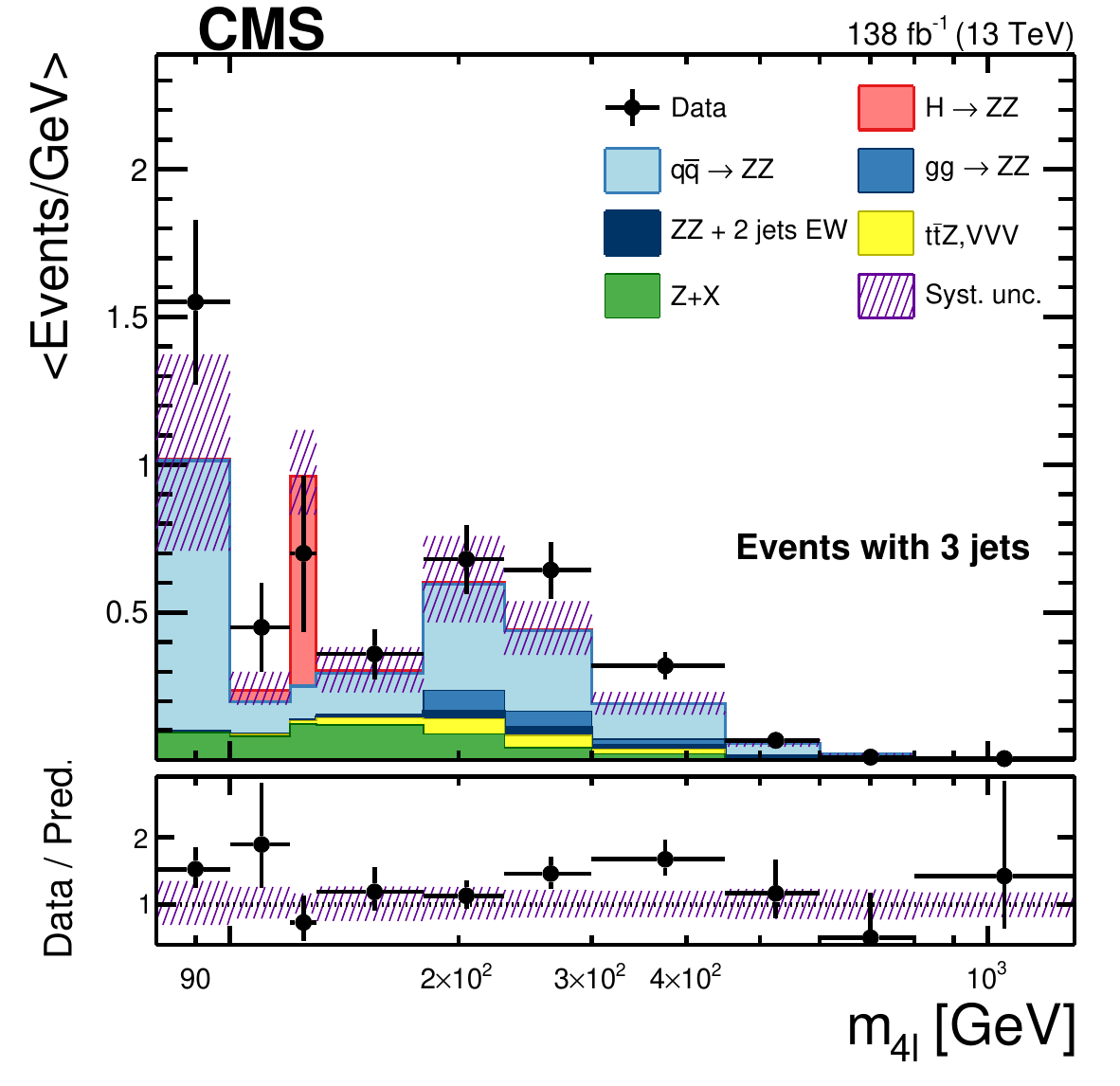}
\includegraphics[width=0.45\textwidth]{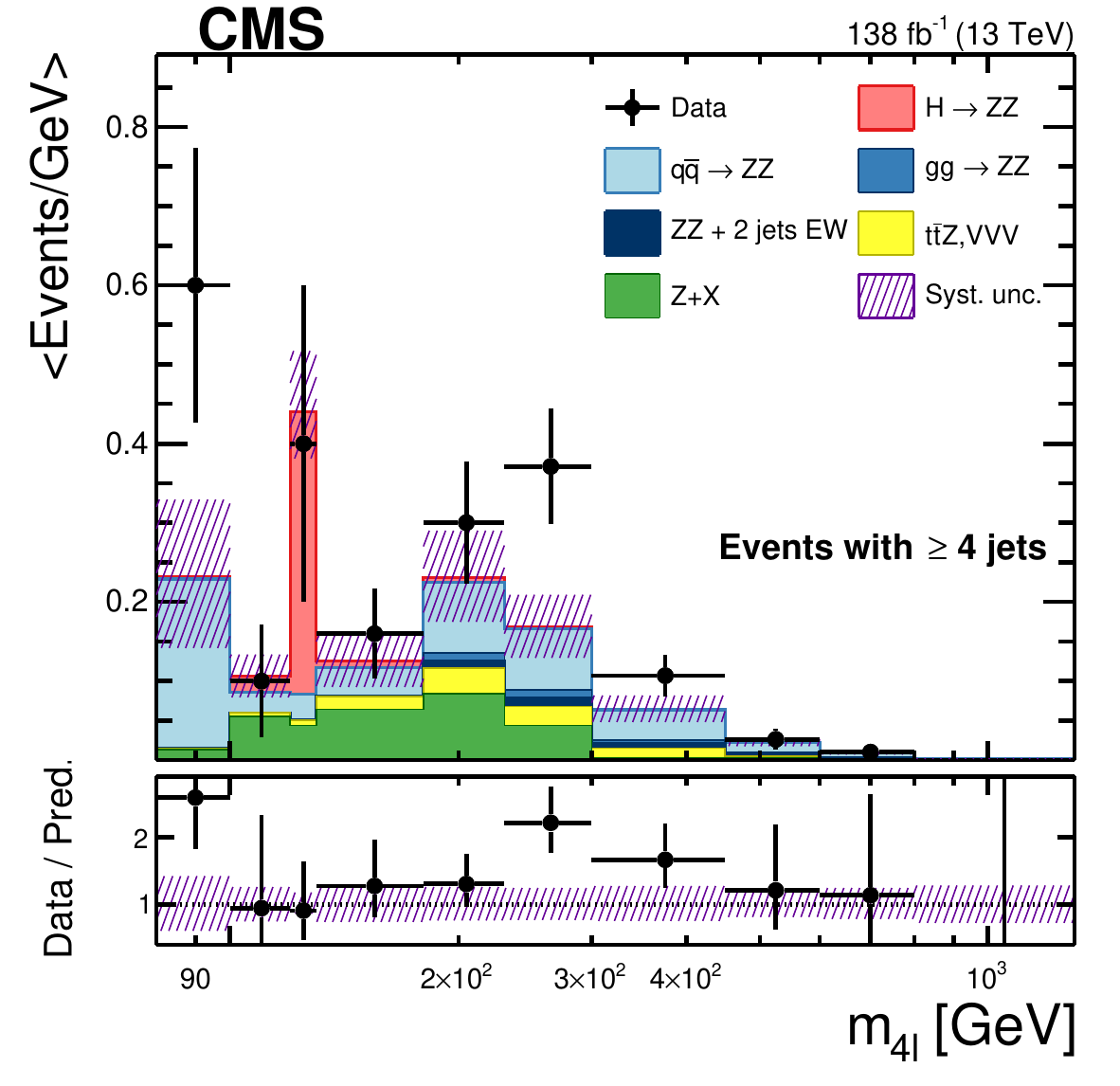}
\caption{
The $m_{4\ell}$ distributions in the full four-lepton invariant mass range for events with different jet multiplicities, normalized by bin width.
Other details are as in the caption of Fig.~\ref{fig:RECO1}.
\label{fig:RECO5}
}
\end{figure}

The effect of the presence of jets in ZZ events is also studied using 
the $m_{4\ell}$ distribution for different jet multiplicities  (Figs.~\ref{fig:RECO4},~\ref{fig:RECO5}). Each distribution  
contains only events with exactly 0, 1, 2, 3 jets or $\geq$4 jets. The predictions describe well the normalized differential behavior, but fail to describe
the event yield in the 1-jet case. With increasing jet multiplicities the predicted yields
decrease much faster than the measured ones. In the case of 4 or more jets, the 
data yields are significantly larger than predicted.
All distributions in Fig.~\ref{fig:RECO4}
are presented for events with on-shell Z bosons, $60 < m_{\PZ_1, \PZ_2} < 120\GeV$.

The same analysis is repeated in the full $m_{4\ell}$ range and the results are presented in Fig.~\ref{fig:RECO5}.
The data and MC predictions are compared in three mass regions:
Z boson region, Higgs boson region, and nonresonant ZZ production region. 
It is important to note that the Higgs boson sample is simulated using the
\POWHEG NLO predictions, whereas a similar contribution in the ZZ sample has the 
$\Pg\Pg \! \to \! \ZZ$ process simulated at LO and normalized
to NLO prediction (see Section~\ref{sec:datasets} for detail).
As shown in Fig.~\ref{fig:RECO5} (upper left), the predictions describe well the data that are inclusive
in jet multiplicity. In Fig.~\ref{fig:RECO5} (middle left), the predictions do not describe the event yield
of the ZZ nonresonant part, but agree well with data in the Z and Higgs boson production regions. 
With increasing jet multiplicity, the agreement between data and predictions
 for ZZ and Z production regions becomes worse, whereas the predictions for the Higgs boson region are compatible with the data within large statistical uncertainties.

The measured and expected event yields for all decay channels and jet multiplicities in different mass ranges
are summarized in Tables~\ref{table:resultsByChannel_Run2} and~\ref{table:resultsByJetMul_Run2}.

\begin{table}[htbp]
  \centering
  \topcaption{The observed and expected yields of $\ZZ$ events in different mass ranges, and estimated yields of background events, shown for each final state and for the sum. The first uncertainty is statistical, and the second one is systematic. (Due to rounding, the sum of individual entries may not match the total value shown.)   }
  \begin{tabular}{ lcccc } 
  \hline
         Process       & $\Pe\Pe\Pe\Pe$  & $\Pe\Pe\PGm\PGm$  & $\PGm\PGm\PGm\PGm$ & $2\ell2\ell'$  \\
  \hline
  \multicolumn{5}{c}{$80< m_{4\ell} < 100\GeV$} \\
  Background          &  $4.6 \pm 0.5 \pm 1.8$   &  $15.5 \pm 1.6 \pm 6.2$  &  $22.8 \pm 2.1 \pm 9.1$  &  $43 \pm 3 \pm 17$	     \\
  Signal              &  $216 \pm 1^{+40}_{-36}$ &  $731 \pm 2^{+66}_{-64}$ &  $841 \pm 2^{+59}_{-57}$ &  $1790 \pm 3^{+140}_{-140}$ \\
  Total expected      &  $220 \pm 1^{+40}_{-36}$ &  $747 \pm 3^{+66}_{-64}$ &  $864 \pm 3^{+59}_{-58}$ &  $1830 \pm 4^{+140}_{-140}$ \\
  Data                & 	   194  	 &  698 		    &  838		       &  1730  		     \\
  \hline
  \multicolumn{5}{c}{$60<m_{\PZ_1,\PZ_2}<120\GeV$} \\
  Background          & $22.9 \pm 0.9 \pm 5.7$  &  $46 \pm 2 \pm 10$          &  $28.9 \pm 1.3 \pm 6.5$   &  $98 \pm 2 \pm 23$          \\
  Signal              & $716 \pm 2^{+63}_{-60}$ &  $1830 \pm 3^{+140}_{-140}$ &  $1138 \pm 3^{+85}_{-82}$ &  $3680 \pm 5^{+280}_{-270}$ \\
  Total expected      & $739 \pm 2^{+63}_{-60}$ &  $1870 \pm 4^{+140}_{-140}$ &  $1167 \pm 3^{+85}_{-82}$ &  $3780 \pm 5^{+280}_{-270}$ \\
  Data                &  671                    &  1805                       &  1106                     &  3582                       \\
  \hline
  \end{tabular}
  \label{table:resultsByChannel_Run2}
  \end{table}

  \begin{table}[htbp]
    \centering
    \topcaption{The observed and expected yields of $\ZZ$ events in different mass ranges, and estimated yields of background events, shown for each jet multiplicity. The first uncertainty is statistical, and the second one is systematic. (Due to rounding, the sum of individual entries may not match the total value shown.)    }
    \resizebox{\columnwidth}{!}{
    \begin{tabular}{ lccccc } 
    \hline
           Process       & 0 jet  & 1 jet & 2 jets & 3 jets & $\geq$4 jets \\
    \hline
    \multicolumn{6}{c}{$80< m_{4\ell} < 100\GeV$} \\
    Background           &  $25 \pm 2 \pm 10$          &  $9.1 \pm 1.3 \pm 3.6$   &  $6.1 \pm 1.0 \pm 2.4$   &  $1.9 \pm 0.6 \pm 0.8$        &  $0.4 \pm 0.3 \pm 0.1$       \\
    Signal               &  $1300 \pm 3^{+100}_{-100}$ &  $371 \pm 2^{+48}_{-45}$ &  $95 \pm 1^{+29}_{-28}$  &  $18.7 \pm 0.4^{+7.1}_{-6.2}$ &  $4.5 \pm 0.2^{+1.9}_{-1.8}$ \\
    Total expected       &  $1320 \pm 3^{+100}_{-100}$ & $ 381 \pm 2^{+48}_{-45}$ &  $101 \pm 1^{+29}_{-28}$ &  $20.6 \pm 0.7^{+7.1}_{-6.2}$ &  $4.9 \pm 0.3^{+2.0}_{-1.8}$ \\
    Data                 &  1238                       &  354                     &  95                      &  31                           &  12                          \\
    \hline
    \multicolumn{6}{c}{$60<m_{\PZ_1,\PZ_2}<120\GeV$} \\
    Background           &  $29.3 \pm 1.4 \pm 8.9$     &  $28.6 \pm 1.2 \pm 6.7$    &  $21.2 \pm 0.9 \pm 3.7$  &  $11.6 \pm 0.7 \pm 2.0$ &  $7.6 \pm 0.5 \pm 1.5$        \\
    Signal               &  $2320 \pm 3^{+160}_{-170}$ &  $960 \pm 3^{+100}_{-90}$  &  $303 \pm 1^{+60}_{-56}$ &  $75 \pm 1^{+20}_{-19}$ &  $21.9 \pm 0.3^{+7.9}_{-7.2}$ \\
    Total expected       &  $2350 \pm 4^{+160}_{-170}$ &  $990 \pm 3^{+100}_{-100}$ &  $324 \pm 2^{+60}_{-56}$ &  $87 \pm 1^{+21}_{-19}$ &  $29.5 \pm 0.7^{+8.1}_{-7.4}$ \\
    Data                 &  2367                       &  741                       &  312                     &  110                    &  52                           \\
    \hline
    \end{tabular}}
    \label{table:resultsByJetMul_Run2}
    \end{table}

\newpage

\subsection{Differential cross sections}
\label{section:diffxsec}

\begin{figure*}[htbp]
\centering
\includegraphics[width=0.48\textwidth]{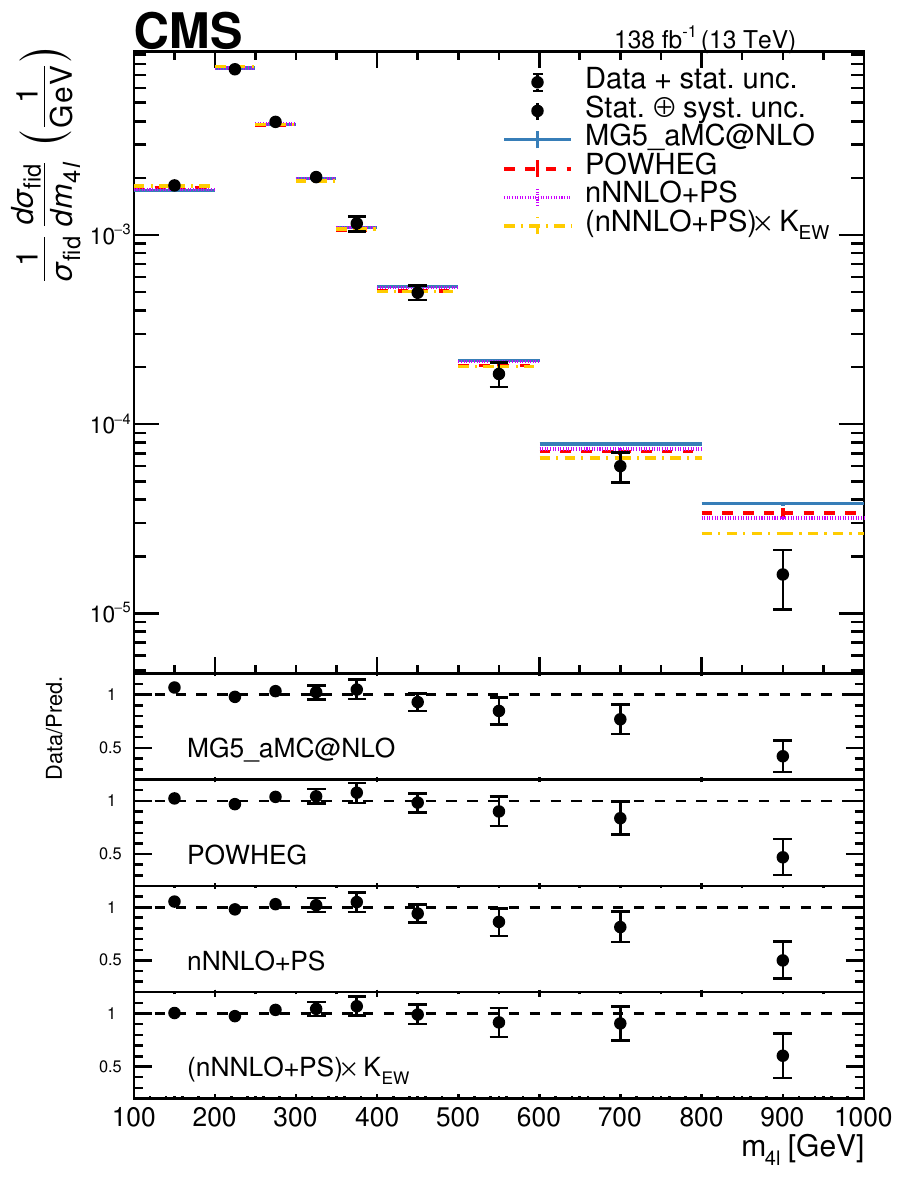}
\includegraphics[width=0.48\textwidth]{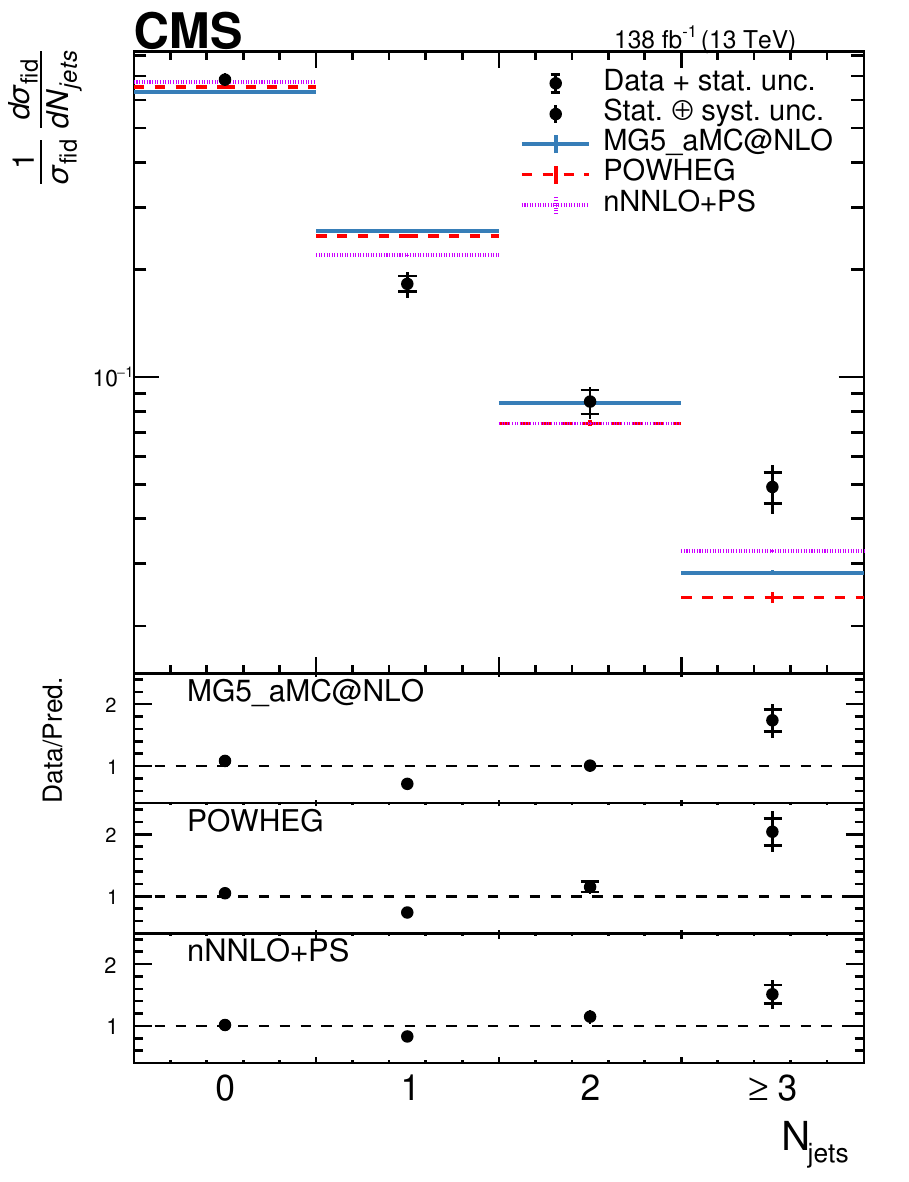}
\caption{Differential cross sections normalized to the fiducial cross section
 as a function of
(left) $m_{4\ell}$, (right) the number of jets with $\pt > 30\GeV$. The on-shell $\PZ$ requirement $60 < m_{\PZ_1, \PZ_2} < 120\GeV$ is applied.
Points represent the unfolded data, solid histograms the 
\MGvATNLO $\qq \to \ZZ$ predictions,
and red dashed histograms
the \POWHEG $\qq \to \ZZ$ predictions. {\MCFM} $\Pg\Pg \to \ZZ$, \POWHEG $\PH \to \ZZ$, and \MGvATNLO EW ZZ predictions are included in these two sets of predictions.  
The purple dashed histograms represent the nNNLO+PS predictions, and the yellow dashed histogram represents the nNNLO+PS prediction with EW corrections applied. 
Vertical bars on both MC predictions represent the statistical uncertainties. 
The lower panels show the ratio of the measured 
to the predicted cross sections.
The vertical bars on data points with horizontal lines at the ends represent the statistical
uncertainties only, whereas the vertical bars without horizontal lines at the ends represent the total uncertainties calculated as the sum in quadrature 
of the statistical and systematic uncertainties. The overflow is included in the last bin of the distributions.
}
\label{fig:diff1}
\end{figure*}

\begin{figure*}[htbp]
\centering
\includegraphics[width=0.48\textwidth]{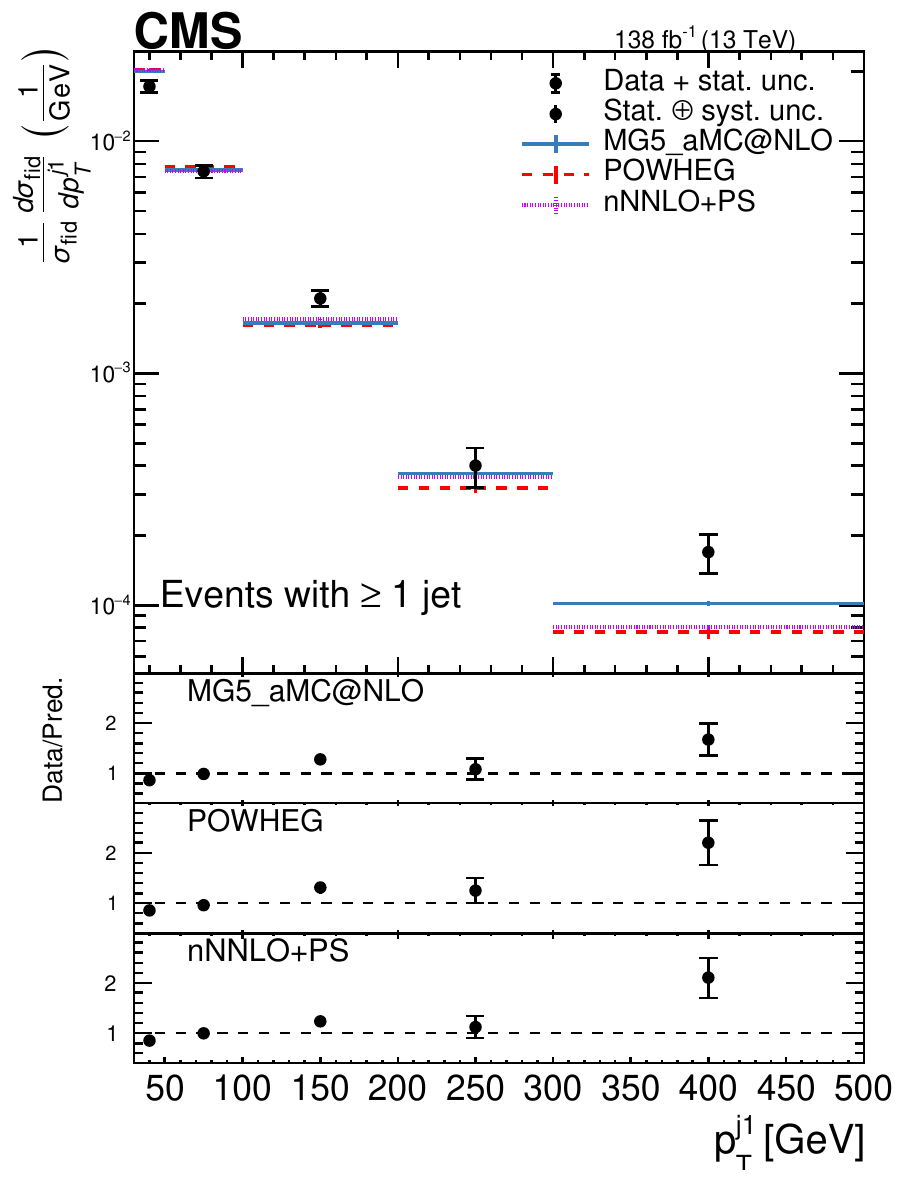}
\includegraphics[width=0.48\textwidth]{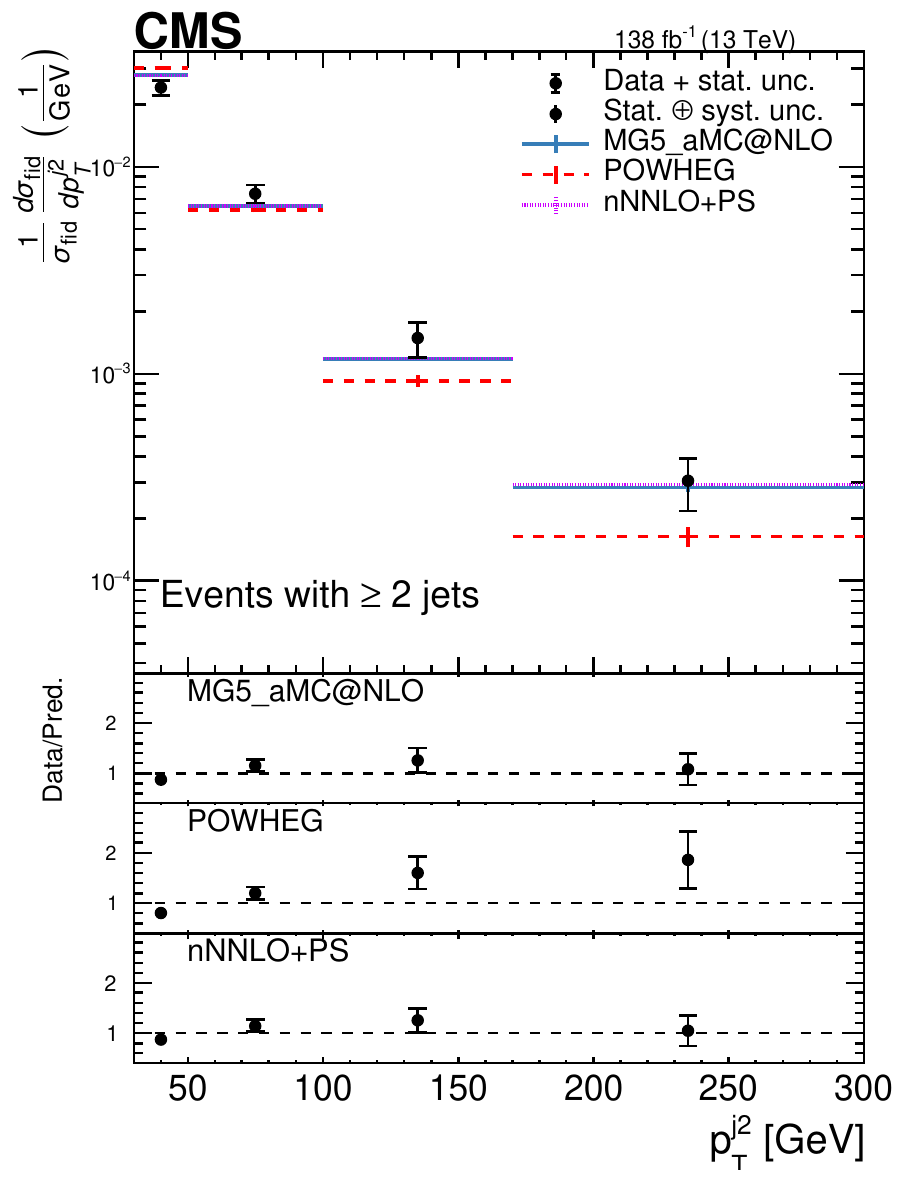} \\
\includegraphics[width=0.48\textwidth]{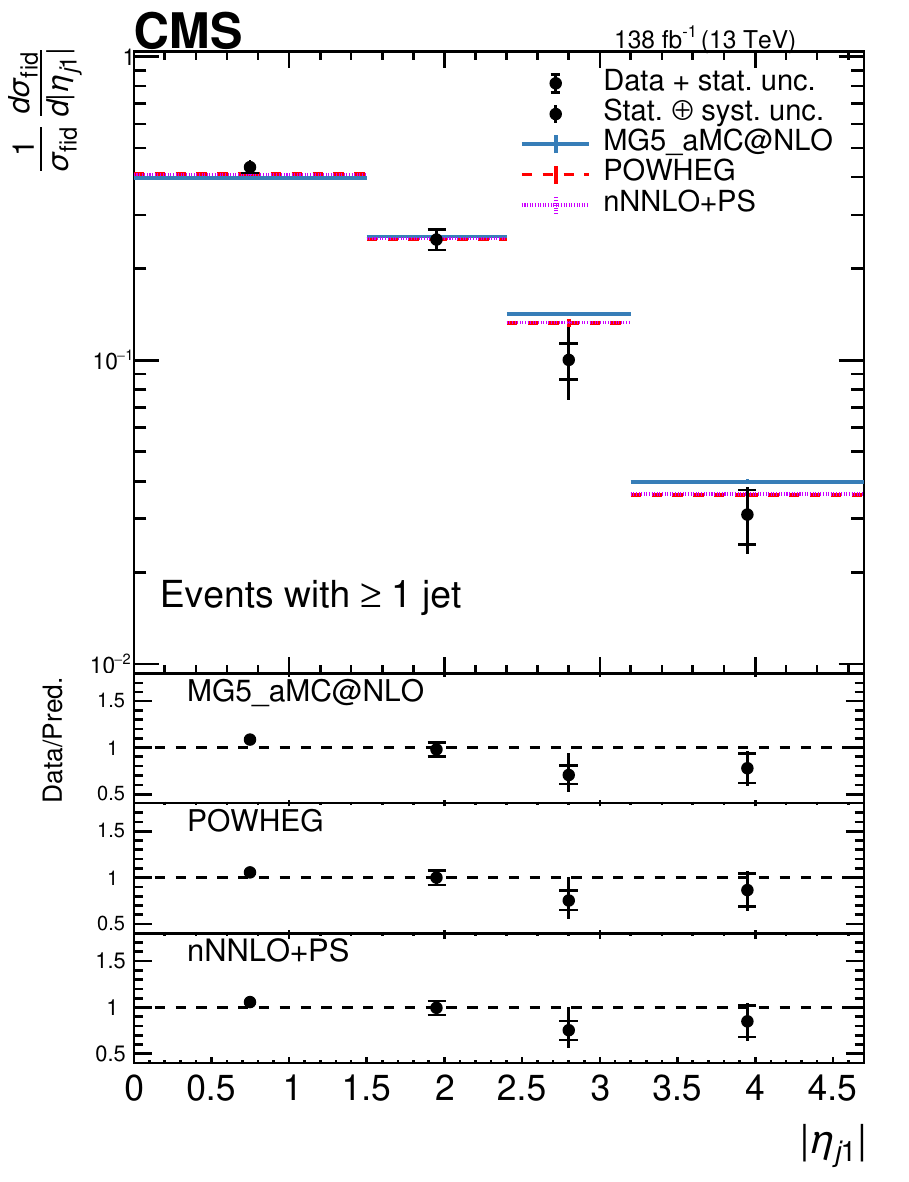}
\includegraphics[width=0.48\textwidth]{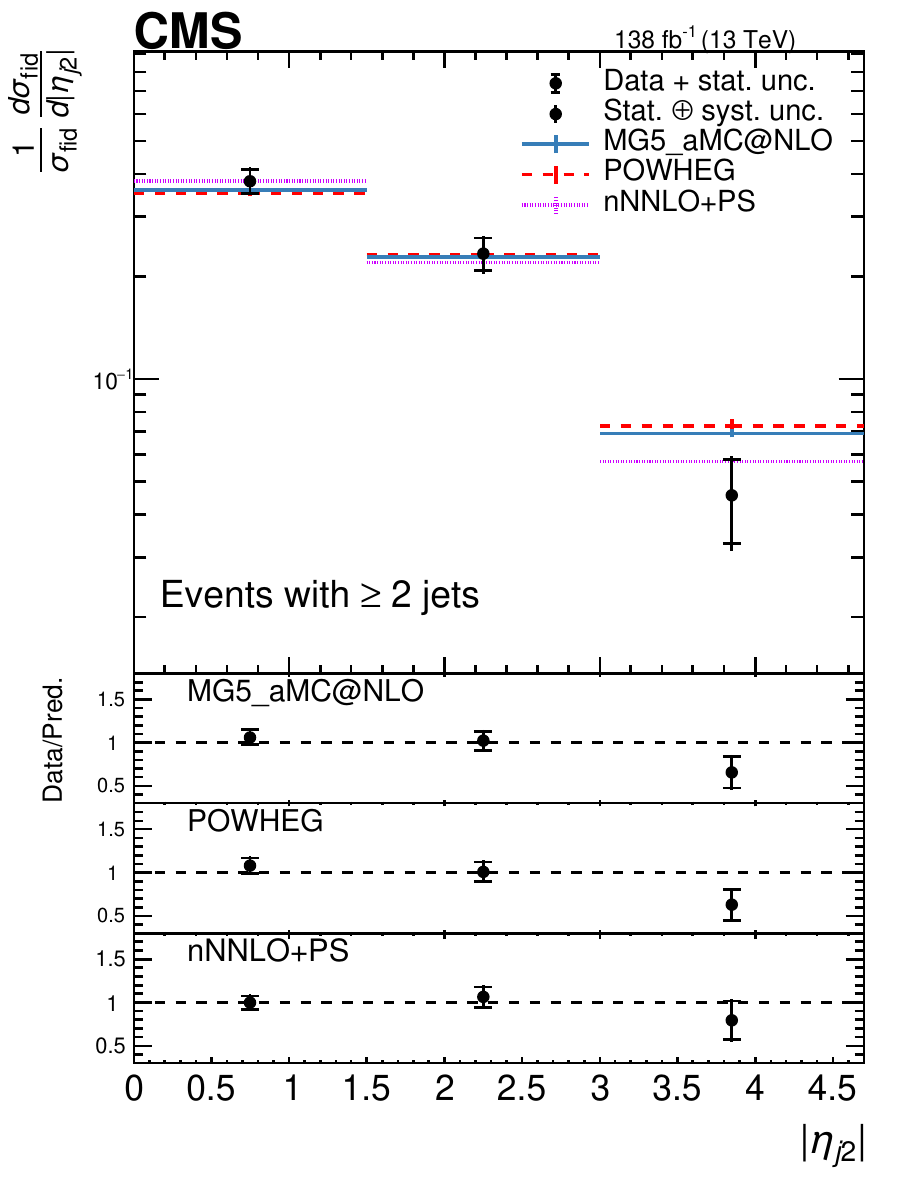}
\caption{Differential cross sections normalized to the fiducial cross section
 as a function of \pt and $\abs{\eta}$ of
 the highest- and the second-highest-\pt jet in events containing at least one or two jets, respectively. The on-shell $\PZ$ requirement $60 < m_{\PZ_1, \PZ_2} < 120\GeV$ is applied.
Other details are as in the caption of Fig.~\ref{fig:diff1}.}
\label{fig:diff2}
\end{figure*}

\begin{figure*}[htbp]
\centering
\includegraphics[width=0.48\textwidth]{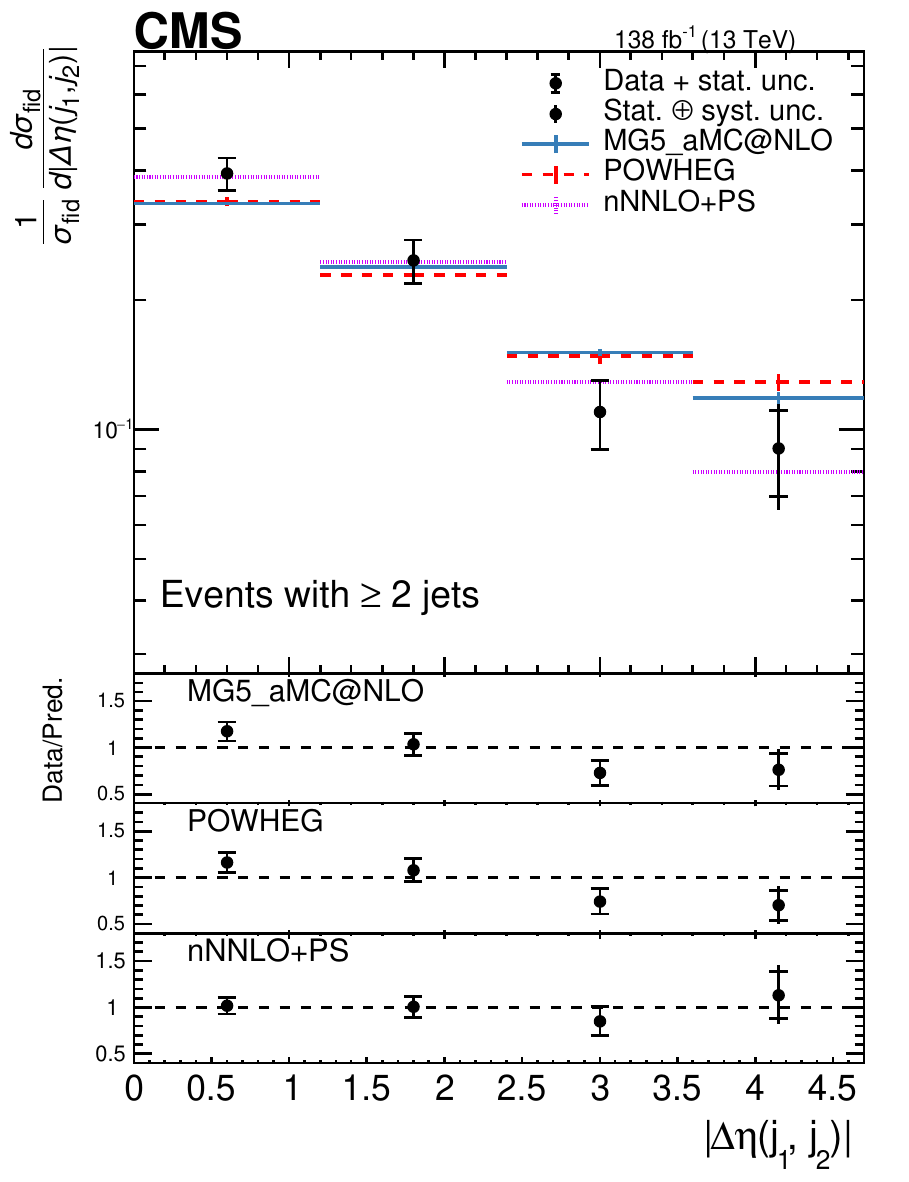}
\includegraphics[width=0.48\textwidth]{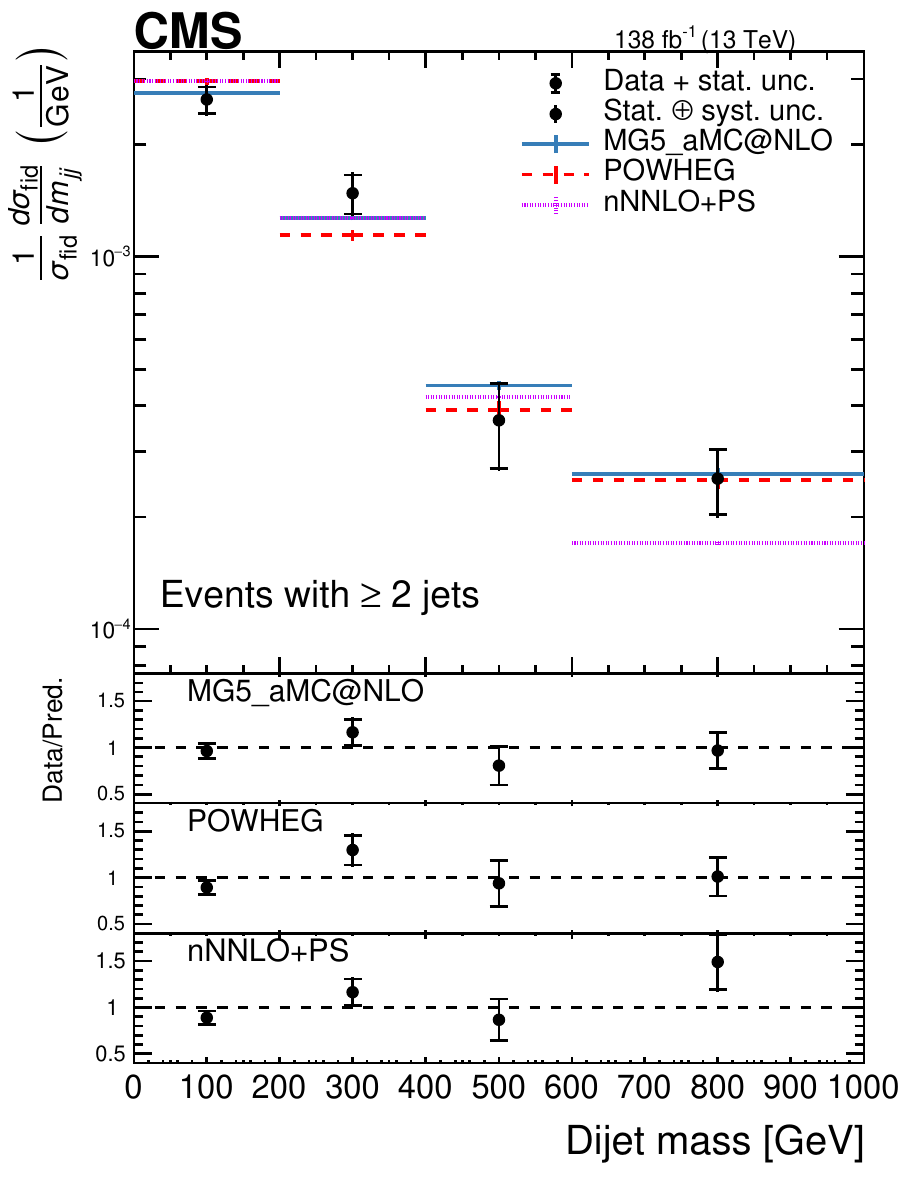}
\caption{Differential cross sections normalized to the fiducial cross section
 as a function of
(left) $\abs{\Delta \eta}$ and (right) dijet mass between highest-\pt jets in events with at least two jets. Events with $60 < m_{\PZ_1, \PZ_2} < 120\GeV$ requirement.
Other details are as in the caption of Fig.~\ref{fig:diff1}.
}
\label{fig:diff3}
\end{figure*}

\begin{figure*}[htbp]
\centering
\includegraphics[width=0.65\textwidth]{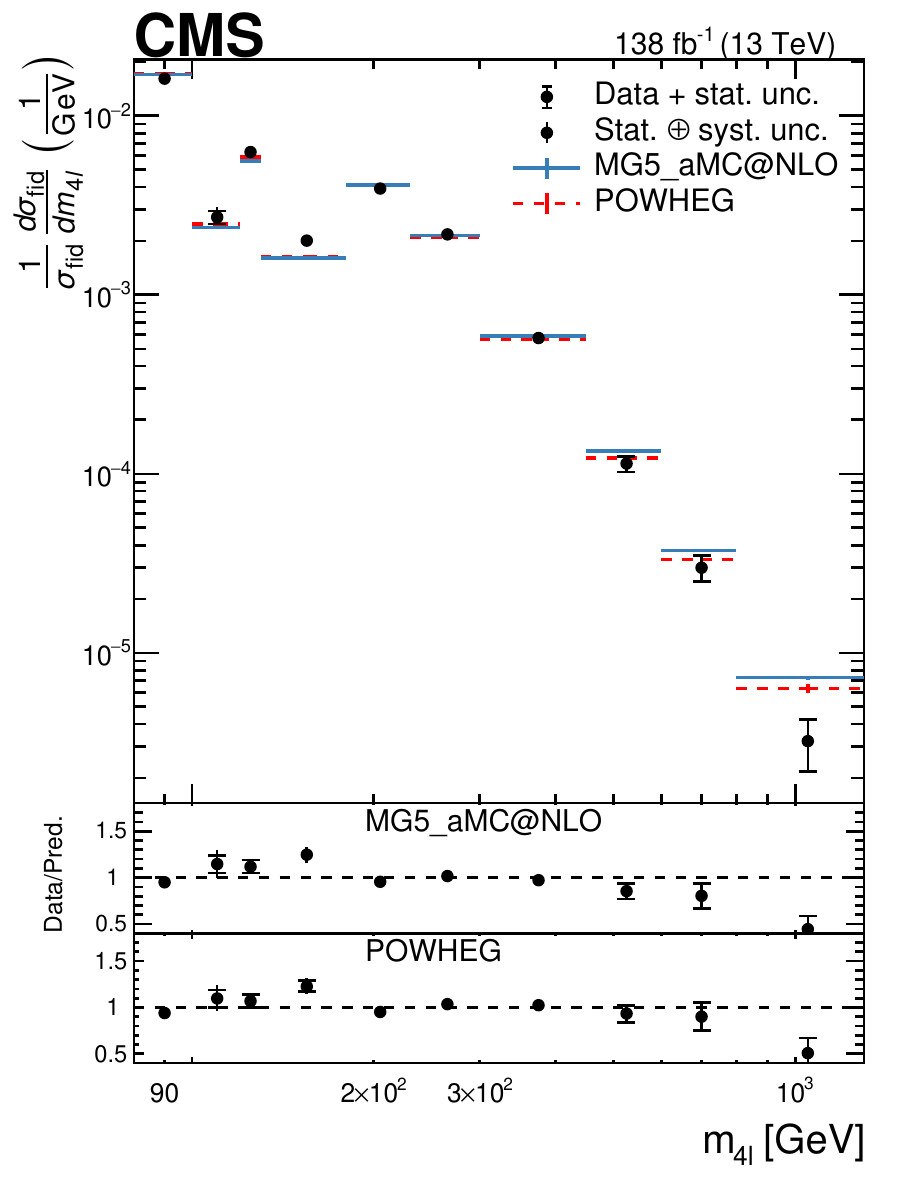}
\caption{Differential cross sections normalized to the fiducial cross section
 as a function of
 $m_{4\ell}$ for the full four-lepton invariant mass range.
Other details are as in the caption of Fig.~\ref{fig:diff1}.
}
\label{fig:diff4}
\end{figure*}

The unfolded differential distributions normalized to the $\ZZ$ fiducial cross section are
presented
in Figs.~\ref{fig:diff1}--\ref{fig:diff4}.
Figure~\ref{fig:diff1} (left) shows the normalized
$\rd\sigma/\rd{m_{4\ell}}$
cross section. The \MGvATNLO, \POWHEG and nNNLO+PS predictions demonstrate
similar behavior and describe well the differential behavior at low $m_{4\ell}$, whereas they
overestimate the measured values in the moderate to high $m_{4\ell}$ regions. This discrepancy can be mitigated with EW corrections as discussed in Ref.~\cite{Buonocore:2021fnj}.
To estimate the effect of the corrections, a differential K factor from~\cite{Buonocore:2021fnj} for the NLO EW corrections was applied to the nNNLO+PS predictions as a function of $m_{4\ell}$. The EW-corrected nNNLO+PS predictions describe the measured values better than those without 
the corrections, although at high values of $m_{4\ell}$ only within large statistical uncertainties of the measurements. The EW corrections are only significant in $m_{4\ell}$ and have negligible effect on any other normalized distribution presented in this paper;
therefore in all other non-$m_{4\ell}$ distributions only nominal nNNLO+PS predictions are shown. For the $m_{4\ell}$ distributions for various jet multiplicities the EW corrections are not available and therefore these distributions do not contain EW corrections.

Figure~\ref{fig:diff1} (right) shows the differential cross sections as a function of the number of jets in the events. 
The \MGvATNLO and \POWHEG predictions show similar distributions. Similar to the discussion in the previous section,
neither of the two MC simulations describes the 1-jet cross section, and both simulations predict too low cross sections for high jet multiplicities.
On the other hand, the nNNLO+PS prediction describes the high jet multiplicity bin better than the other two predictions, whereas the agreement in the 1-jet bin is also improved. In general, the nNNLO+PS prediction describes the N\textsubscript{jets} distribution better than \MGvATNLO and \POWHEG.

Figure~\ref{fig:diff2} shows the differential cross sections in bins of \pt and $\abs{\eta}$ for
the highest- and the second highest-\pt jet in events with at least one and two jets.
The \pt distributions show moderate differences between
data and predictions, whereas the $\abs{\eta}$ distributions are well described
within uncertainties.

Figure~\ref{fig:diff3} shows the differential cross sections for dijet events as a function of (left) $\abs{\Delta \eta}$ and 
(right) the dijet mass between two highest-\pt jets. 
Within uncertainties the dijet mass is described by the predictions, whereas $\abs{\Delta \eta}$ measurements show
a small trend with respect to the predictions.

\begin{figure*}[htbp]
\centering
\includegraphics[width=0.48\textwidth]{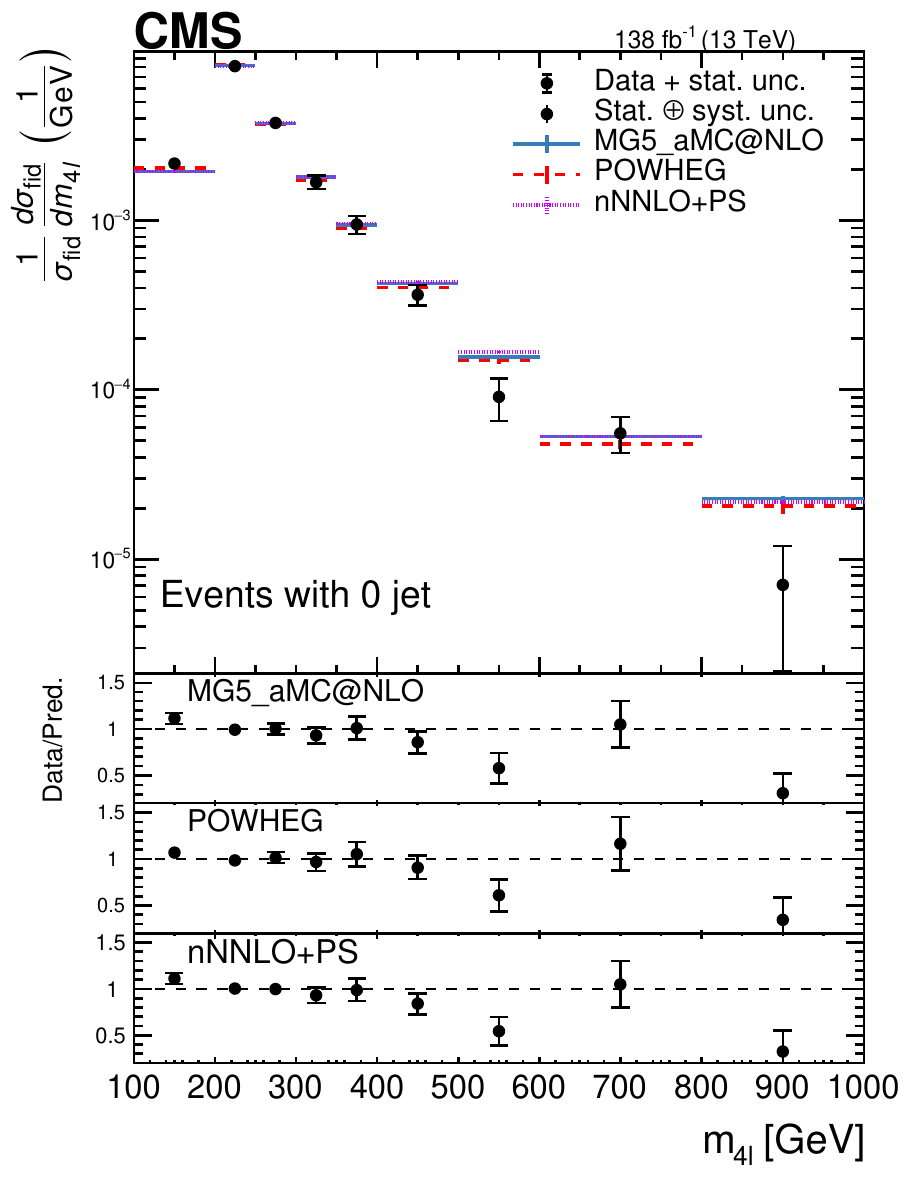}
\includegraphics[width=0.48\textwidth]{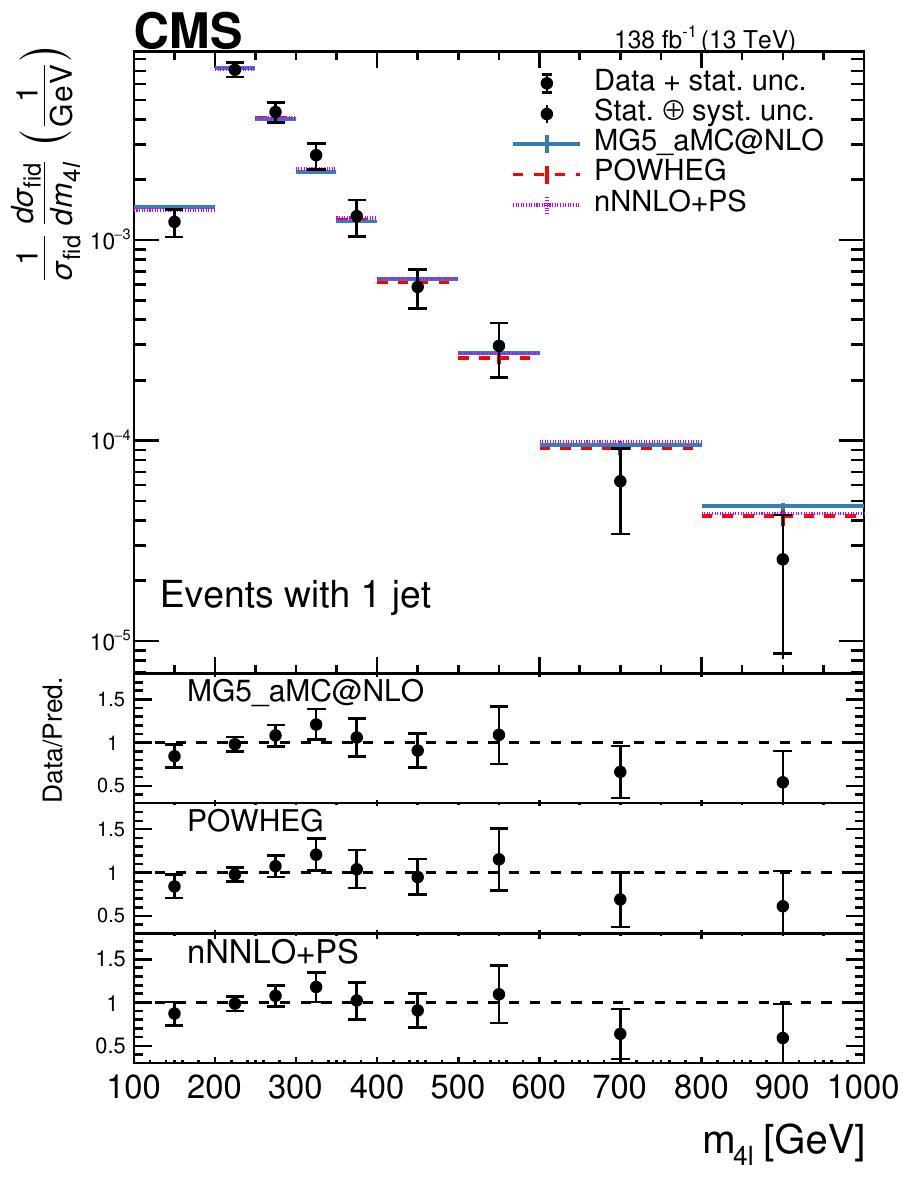} \\
\includegraphics[width=0.48\textwidth]{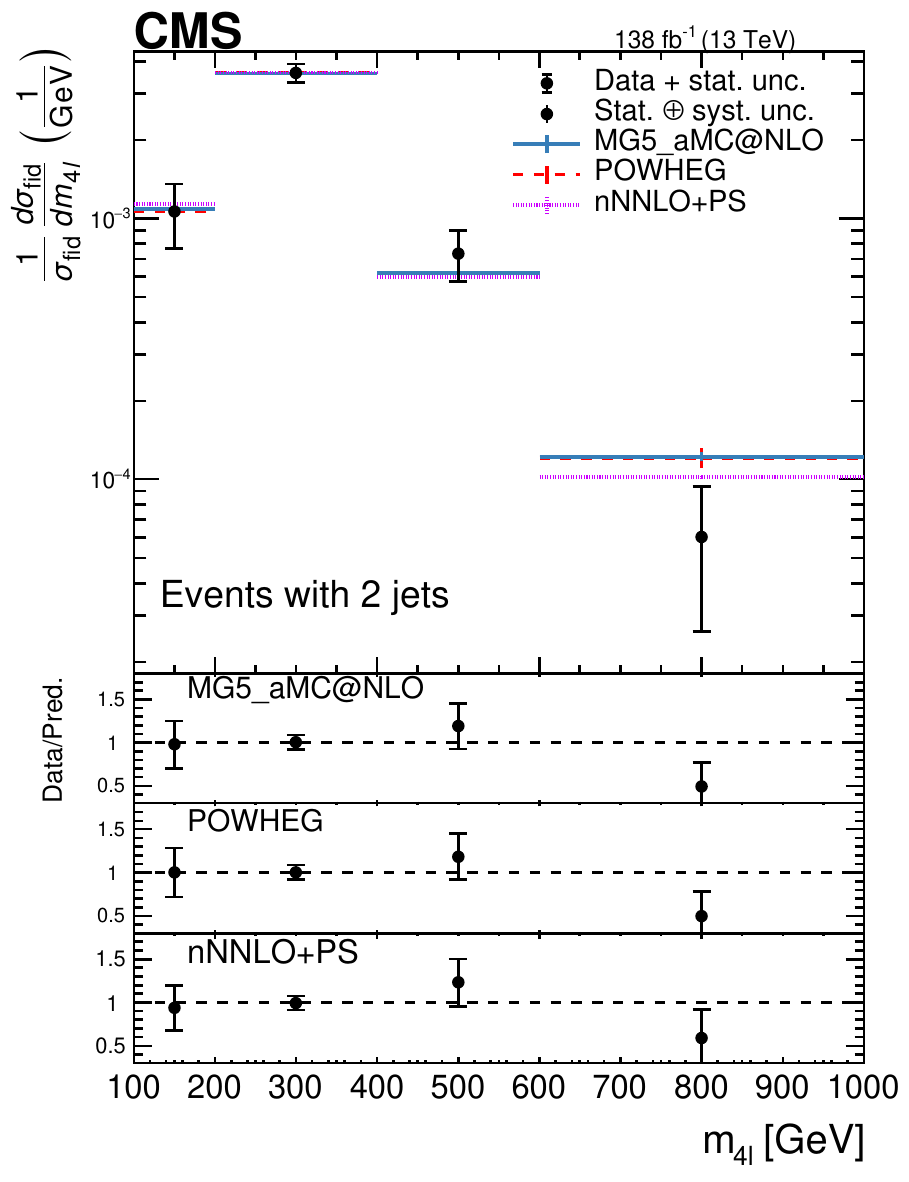}
\includegraphics[width=0.48\textwidth]{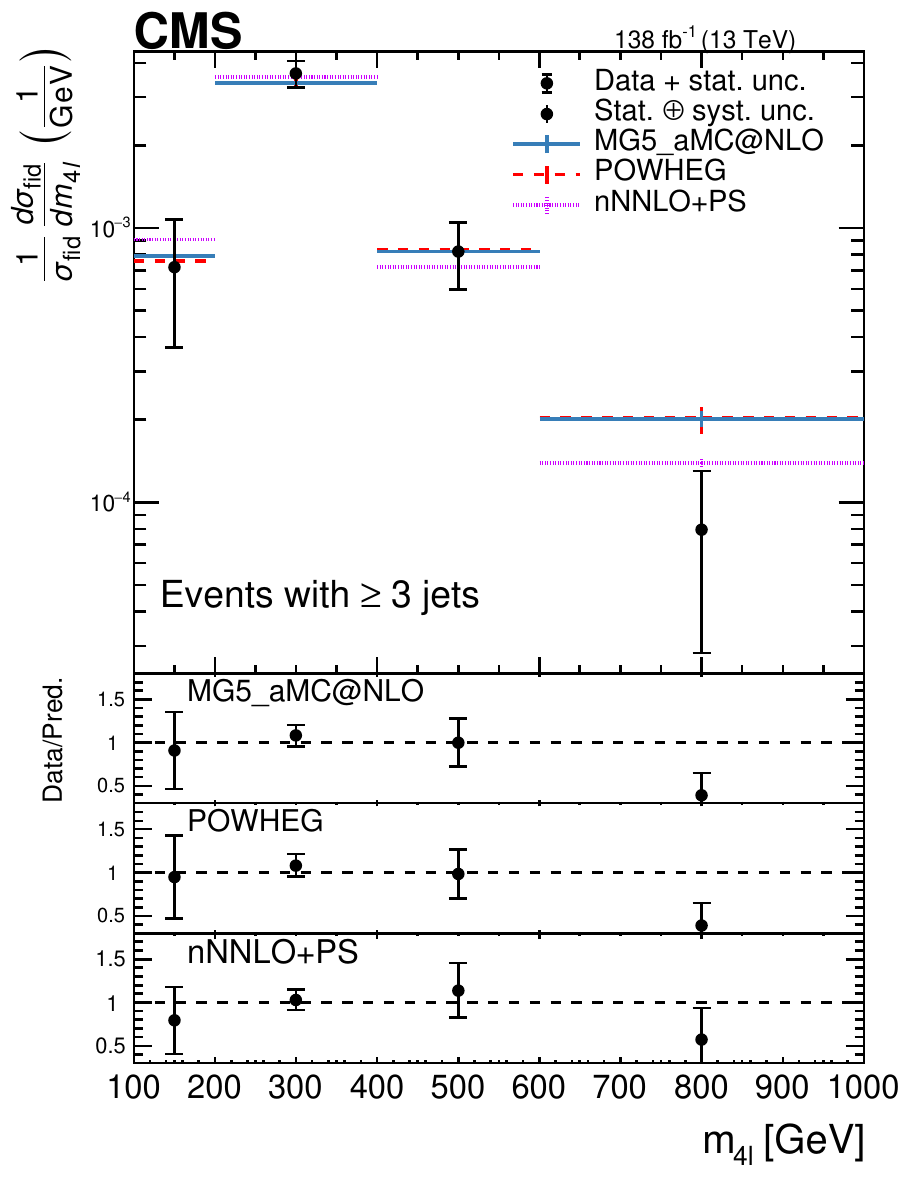}
\caption{Differential cross sections normalized to the fiducial cross section
 as a function of
 $m_{4\ell}$ for $60 < m_{\PZ_1, \PZ_2} < 120\GeV$ and for different jet multiplicities. Other details are as in the Fig.~\ref{fig:diff1} caption.
}
\label{fig:diffa1}
\end{figure*}

\begin{figure*}[htbp]
\centering
\includegraphics[width=0.48\textwidth]{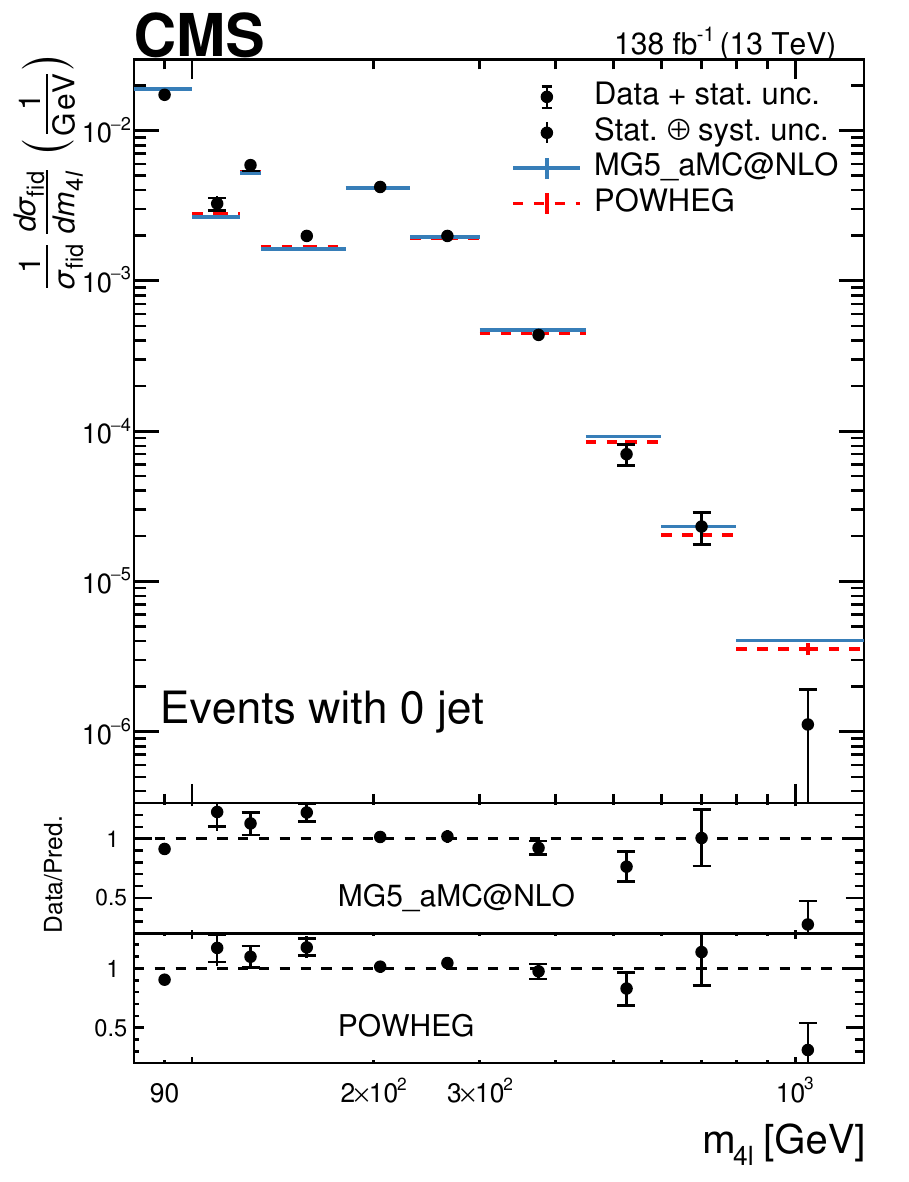}
\includegraphics[width=0.48\textwidth]{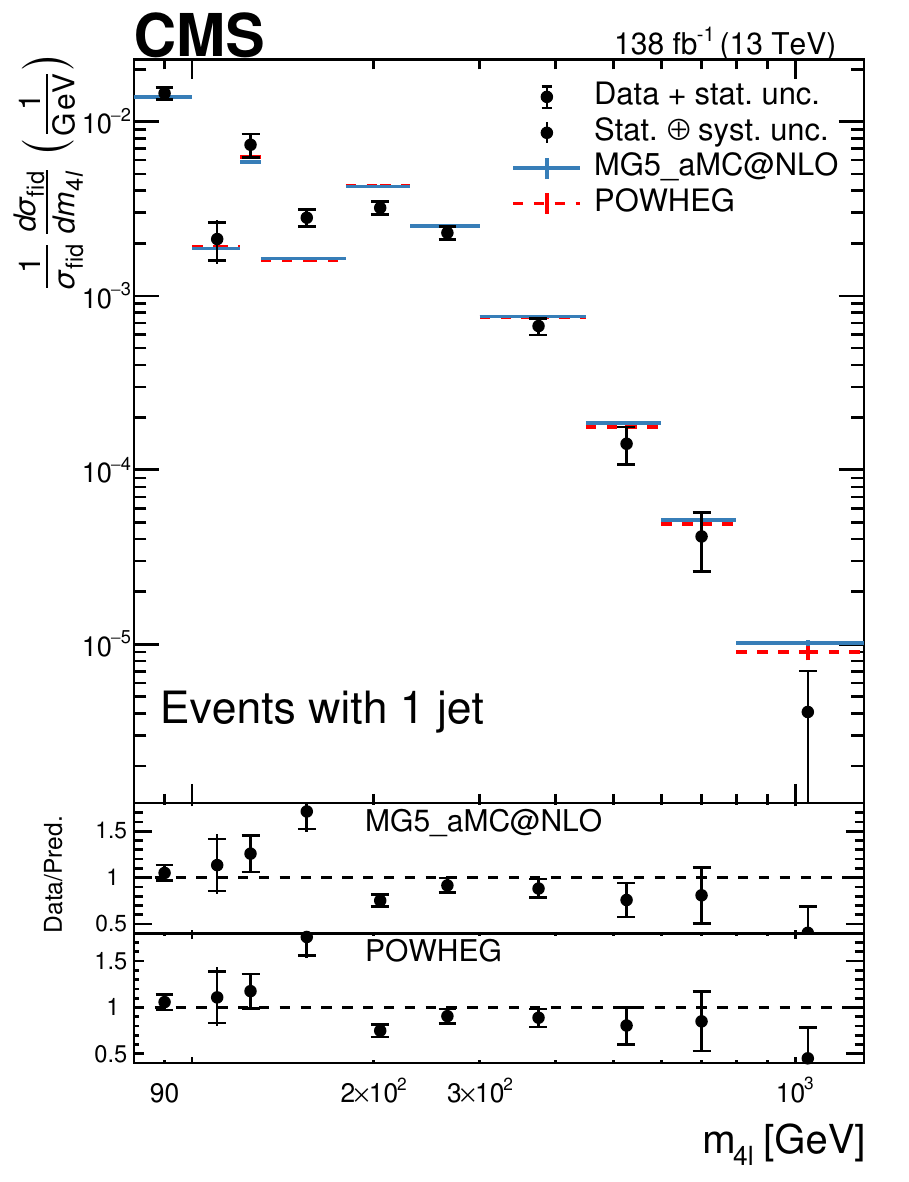}
\includegraphics[width=0.48\textwidth]{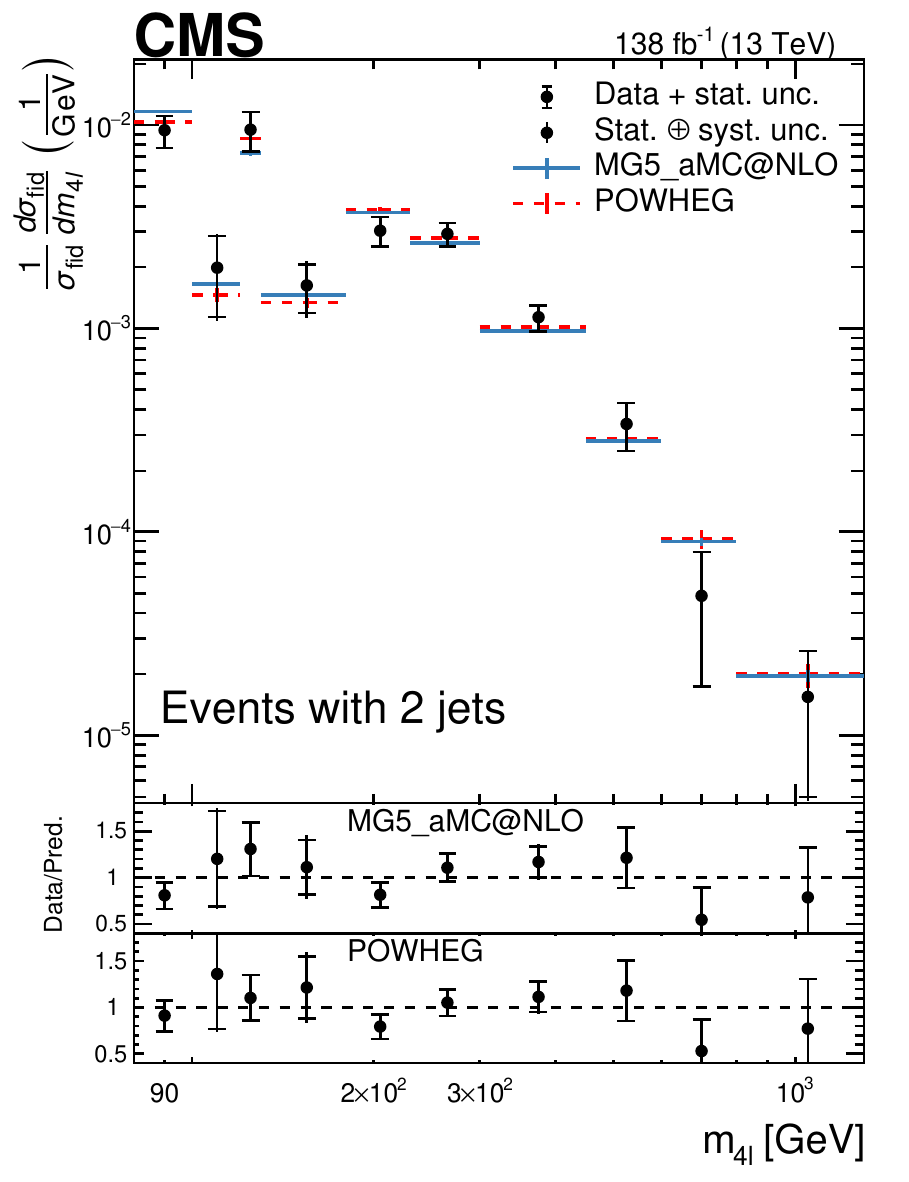}
\includegraphics[width=0.48\textwidth]{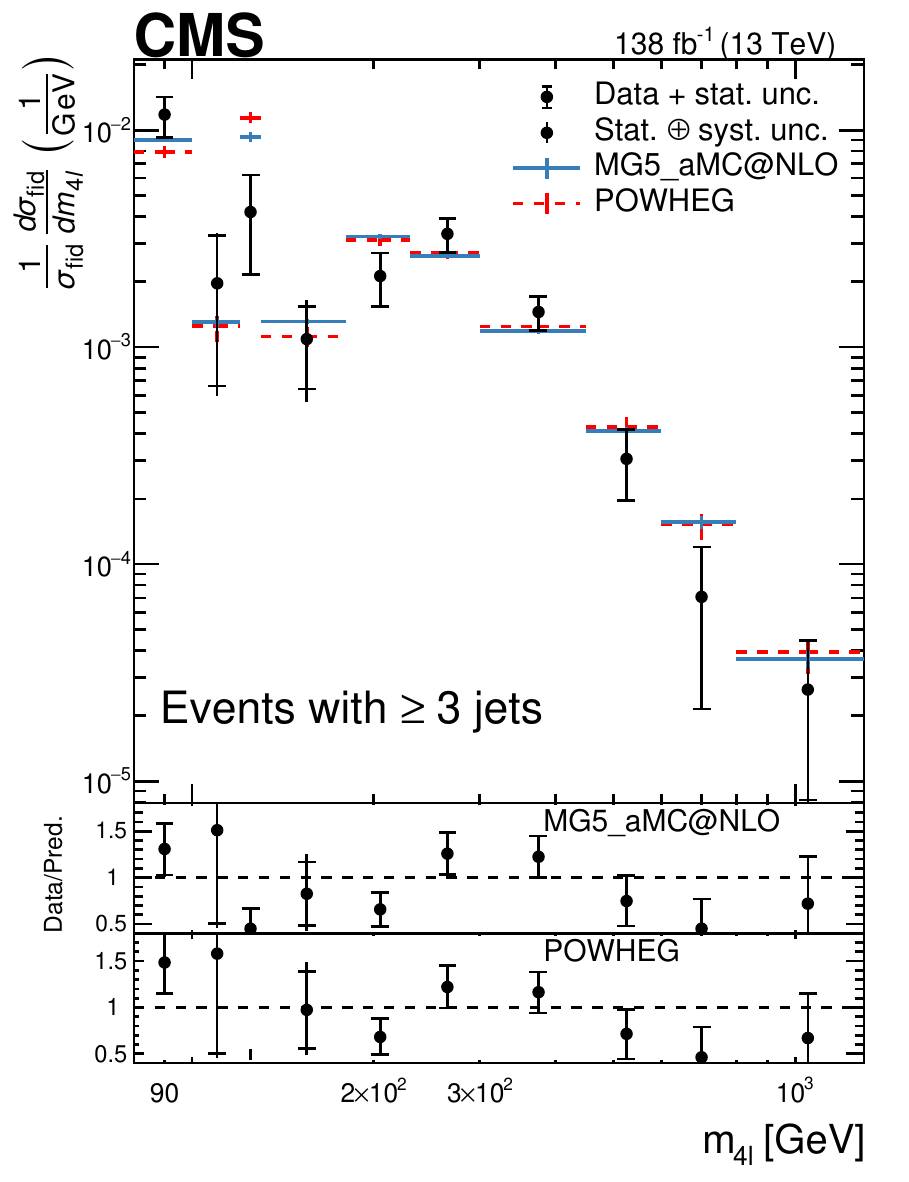}
\caption{Differential cross sections normalized to the fiducial cross section
as a function of
$m_{4\ell}$ for the full four-lepton invariant mass range and different jet multiplicities. Other details are as in the Fig.~\ref{fig:diff1} caption.
}
\label{fig:diffa2}
\end{figure*}

Finally, the $\rd\sigma/\rd{m_{4\ell}}$ differential cross section is presented in Fig.~\ref{fig:diff4} 
for the full four-lepton invariant mass range and inclusive in jet multiplicity. The measured normalized differential cross section is well described by the predictions. 
Additional $\rd\sigma/\rd{m_{4\ell}}$ differential cross sections with different jet multiplicities for the on-shell Z bosons and for the full four-lepton invariant mass range
are presented in Figs.~\ref{fig:diffa1} and~\ref{fig:diffa2}. 
The comparison with the theoretical predictions shows
the same behavior than for the measurements presented in the previous section.

\section{Summary}

The four-lepton production in association with jets, $\pp \to (\PZ/\PGg^*)(\PZ/\PGg^*) +\text{jets} \to 2\ell2\ell'+\text{jets}$, where $\ell,\ell' = \Pe$ or $\PGm$,  
was studied in proton-proton collisions at a center-of-mass energy of 13\TeV. The data sample corresponds to an integrated luminosity of 138\fbinv collected with the CMS detector at the LHC during 2016--2018.
Differential distributions and differential cross sections normalized to the $\ZZ$ fiducial cross section were measured 
with respect to various kinematic variables: 
number of jets, jet transverse momentum (\pt) and pseudorapidity ($\eta$), invariant mass of the dijet system and $\eta$ difference between the highest-\pt and second-highest-\pt jets, and invariant mass of the four leptons ($m_{4\ell}$) for different jet multiplicities. Tabulated results are provided in HEPData~\cite{HEPData}.
In general, predictions of theoretical models agree with the data, but in some regions significant discrepancies
between predicted and measured values were observed. The recent nNNLO+PS prediction improves the data/prediction agreement in the 1-jet and high jet multiplicity regions, and 
describes the distribution of jet multiplicities better than NLO samples generated with the event generators 
\MGvATNLO and \POWHEG. The inclusion of electroweak corrections improves the description of the $m_{4\ell}$ distribution. These measurements demonstrate 
the necessity for better Monte Carlo modeling in events with complex multiboson final states and extra jets. 
Further improvement of the predictions is required to
describe the $\ZZ$+jets production in the whole phase space.

\begin{acknowledgments}
  We congratulate our colleagues in the CERN accelerator departments for the excellent performance of the LHC and thank the technical and administrative staffs at CERN and at other CMS institutes for their contributions to the success of the CMS effort. In addition, we gratefully acknowledge the computing centers and personnel of the Worldwide LHC Computing Grid and other centers for delivering so effectively the computing infrastructure essential to our analyses. Finally, we acknowledge the enduring support for the construction and operation of the LHC, the CMS detector, and the supporting computing infrastructure provided by the following funding agencies: SC (Armenia), BMBWF and FWF (Austria); FNRS and FWO (Belgium); CNPq, CAPES, FAPERJ, FAPERGS, and FAPESP (Brazil); MES and BNSF (Bulgaria); CERN; CAS, MoST, and NSFC (China); MINCIENCIAS (Colombia); MSES and CSF (Croatia); RIF (Cyprus); SENESCYT (Ecuador); ERC PRG, RVTT3 and MoER TK202 (Estonia); Academy of Finland, MEC, and HIP (Finland); CEA and CNRS/IN2P3 (France); SRNSF (Georgia); BMBF, DFG, and HGF (Germany); GSRI (Greece); NKFIH (Hungary); DAE and DST (India); IPM (Iran); SFI (Ireland); INFN (Italy); MSIP and NRF (Republic of Korea); MES (Latvia); LMTLT (Lithuania); MOE and UM (Malaysia); BUAP, CINVESTAV, CONACYT, LNS, SEP, and UASLP-FAI (Mexico); MOS (Montenegro); MBIE (New Zealand); PAEC (Pakistan); MES and NSC (Poland); FCT (Portugal); MESTD (Serbia); MCIN/AEI and PCTI (Spain); MOSTR (Sri Lanka); Swiss Funding Agencies (Switzerland); MST (Taipei); MHESI and NSTDA (Thailand); TUBITAK and TENMAK (Turkey); NASU (Ukraine); STFC (United Kingdom); DOE and NSF (USA).
   
  \hyphenation{Rachada-pisek} Individuals have received support from the Marie-Curie program and the European Research Council and Horizon 2020 Grant, contract Nos.\ 675440, 724704, 752730, 758316, 765710, 824093, 101115353,101002207, and COST Action CA16108 (European Union); the Leventis Foundation; the Alfred P.\ Sloan Foundation; the Alexander von Humboldt Foundation; the Science Committee, project no. 22rl-037 (Armenia); the Belgian Federal Science Policy Office; the Fonds pour la Formation \`a la Recherche dans l'Industrie et dans l'Agriculture (FRIA-Belgium); the Agentschap voor Innovatie door Wetenschap en Technologie (IWT-Belgium); the F.R.S.-FNRS and FWO (Belgium) under the ``Excellence of Science -- EOS" -- be.h project n.\ 30820817; the Beijing Municipal Science \& Technology Commission, No. Z191100007219010 and Fundamental Research Funds for the Central Universities (China); the Ministry of Education, Youth and Sports (MEYS) of the Czech Republic; the Shota Rustaveli National Science Foundation, grant FR-22-985 (Georgia); the Deutsche Forschungsgemeinschaft (DFG), under Germany's Excellence Strategy -- EXC 2121 ``Quantum Universe" -- 390833306, and under project number 400140256 - GRK2497; the Hellenic Foundation for Research and Innovation (HFRI), Project Number 2288 (Greece); the Hungarian Academy of Sciences, the New National Excellence Program - \'UNKP, the NKFIH research grants K 131991, K 133046, K 138136, K 143460, K 143477, K 146913, K 146914, K 147048, 2020-2.2.1-ED-2021-00181, and TKP2021-NKTA-64 (Hungary); the Council of Science and Industrial Research, India; ICSC -- National Research Center for High Performance Computing, Big Data and Quantum Computing, funded by the EU NexGeneration program (Italy); the Latvian Council of Science; the Ministry of Education and Science, project no. 2022/WK/14, and the National Science Center, contracts Opus 2021/41/B/ST2/01369 and 2021/43/B/ST2/01552 (Poland); the Funda\c{c}\~ao para a Ci\^encia e a Tecnologia, grant CEECIND/01334/2018 (Portugal); the National Priorities Research Program by Qatar National Research Fund; MCIN/AEI/10.13039/501100011033, ERDF ``a way of making Europe", and the Programa Estatal de Fomento de la Investigaci{\'o}n Cient{\'i}fica y T{\'e}cnica de Excelencia Mar\'{\i}a de Maeztu, grant MDM-2017-0765 and Programa Severo Ochoa del Principado de Asturias (Spain); the Chulalongkorn Academic into Its 2nd Century Project Advancement Project, and the National Science, Research and Innovation Fund via the Program Management Unit for Human Resources \& Institutional Development, Research and Innovation, grant B37G660013 (Thailand); the Kavli Foundation; the Nvidia Corporation; the SuperMicro Corporation; the Welch Foundation, contract C-1845; and the Weston Havens Foundation (USA).  
\end{acknowledgments}

\bibliography{auto_generated}
\cleardoublepage \appendix\section{The CMS Collaboration \label{app:collab}}\begin{sloppypar}\hyphenpenalty=5000\widowpenalty=500\clubpenalty=5000
\cmsinstitute{Yerevan Physics Institute, Yerevan, Armenia}
{\tolerance=6000
A.~Hayrapetyan, A.~Tumasyan\cmsAuthorMark{1}\cmsorcid{0009-0000-0684-6742}
\par}
\cmsinstitute{Institut f\"{u}r Hochenergiephysik, Vienna, Austria}
{\tolerance=6000
W.~Adam\cmsorcid{0000-0001-9099-4341}, J.W.~Andrejkovic, T.~Bergauer\cmsorcid{0000-0002-5786-0293}, S.~Chatterjee\cmsorcid{0000-0003-2660-0349}, K.~Damanakis\cmsorcid{0000-0001-5389-2872}, M.~Dragicevic\cmsorcid{0000-0003-1967-6783}, A.~Escalante~Del~Valle\cmsorcid{0000-0002-9702-6359}, P.S.~Hussain\cmsorcid{0000-0002-4825-5278}, M.~Jeitler\cmsAuthorMark{2}\cmsorcid{0000-0002-5141-9560}, N.~Krammer\cmsorcid{0000-0002-0548-0985}, D.~Liko\cmsorcid{0000-0002-3380-473X}, I.~Mikulec\cmsorcid{0000-0003-0385-2746}, J.~Schieck\cmsAuthorMark{2}\cmsorcid{0000-0002-1058-8093}, R.~Sch\"{o}fbeck\cmsorcid{0000-0002-2332-8784}, D.~Schwarz\cmsorcid{0000-0002-3821-7331}, M.~Sonawane\cmsorcid{0000-0003-0510-7010}, S.~Templ\cmsorcid{0000-0003-3137-5692}, W.~Waltenberger\cmsorcid{0000-0002-6215-7228}, C.-E.~Wulz\cmsAuthorMark{2}\cmsorcid{0000-0001-9226-5812}
% ...
\par}
\vskip\cmsinstskip
\dag:~Deceased\\
$^{1}$Also at Yerevan State University, Yerevan, Armenia\\
$^{2}$Also at TU Wien, Vienna, Austria\\
% ...
\end{sloppypar}
\end{document}